\documentclass[%
reprint,
superscriptaddress,
amsmath,amssymb,
aps,longbibliography,
pra,
]{revtex4-2}

\usepackage{graphicx}%
\usepackage{verbatim}
\usepackage[version=4]{mhchem}
\usepackage{xcolor}
\usepackage{physics}
\usepackage[normalem]{ulem}
\usepackage{pdfpages}
\usepackage{pgffor}
\usepackage{hyperref}
\usepackage{tabularx}
\usepackage{soul}
\usepackage{booktabs}
\usepackage{enumitem}

\makeatletter
\AtBeginDocument{\let\LS@rot\@undefined}
\makeatother

\usepackage[normalem]{ulem}
\definecolor{tangerine}{rgb}{0.944,0.522,0}
\definecolor{verde}{rgb}{0.,0.6,0}
\definecolor{rosso}{rgb}{0.9,0.0,0.2}
\definecolor{orange}{rgb}{1.0,0.5,0.0}

\newif\ifhighlight

\newcommand{\highlight}{\highlighttrue}
\highlight
\newcommand{\editor}[2]{%
  \expandafter\newcommand\csname #1note\endcsname[1]{%
    \textcolor{#2}{(\textbf{#1note:} \textsc{##1})}}%
  \expandafter\newcommand\csname #1\endcsname[1]{%
    \ifhighlight\textcolor{#2}{##1} \else ##1\fi}%
  \expandafter\newcommand\csname #1cancel\endcsname[1]{%
    \ifhighlight\textcolor{#2}{\sout{##1}}\fi}%
  \expandafter\newcommand\csname #1change\endcsname[2]{%
    \ifhighlight\textcolor{#2}{\sout{##1} ##2}\else ##2\fi}%
  \newenvironment{#1text}{\ifhighlight\color{#2}\fi}{\color{black}}
}

\editor{MC}{red}
\editor{resub}{blue}
\editor{DT}{verde}
\editor{PP}{purple}
\editor{MK}{verde}
\editor{AM}{orange}

\begin{document}

\title{A universal machine learning model for the electronic density of states}

\newcommand{\FB}[1] {{\color{teal} #1}}
\newcommand{\FBcancel}[1] {{\color{teal} \st{#1}}}

\author{Wei Bin How}
\affiliation{Laboratory of Computational Science and Modeling, Institut des Mat\'eriaux, \'Ecole Polytechnique F\'ed\'erale de Lausanne, 1015 Lausanne, Switzerland}

\author{Pol Febrer}
\affiliation{Laboratory of Computational Science and Modeling, Institut des Mat\'eriaux, \'Ecole Polytechnique F\'ed\'erale de Lausanne, 1015 Lausanne, Switzerland}

\author{Sanggyu Chong}
\affiliation{Laboratory of Computational Science and Modeling, Institut des Mat\'eriaux, \'Ecole Polytechnique F\'ed\'erale de Lausanne, 1015 Lausanne, Switzerland}

\author{Arslan Mazitov}
\affiliation{Laboratory of Computational Science and Modeling, Institut des Mat\'eriaux, \'Ecole Polytechnique F\'ed\'erale de Lausanne, 1015 Lausanne, Switzerland}

\author{Filippo Bigi}
\affiliation{Laboratory of Computational Science and Modeling, Institut des Mat\'eriaux, \'Ecole Polytechnique F\'ed\'erale de Lausanne, 1015 Lausanne, Switzerland}

\author{Matthias Kellner}
\affiliation{Laboratory of Computational Science and Modeling, Institut des Mat\'eriaux, \'Ecole Polytechnique F\'ed\'erale de Lausanne, 1015 Lausanne, Switzerland}

\author{Sergey Pozdnyakov}
\affiliation{Laboratory of Computational Science and Modeling, Institut des Mat\'eriaux, \'Ecole Polytechnique F\'ed\'erale de Lausanne, 1015 Lausanne, Switzerland}

\author{Michele Ceriotti}
\email{michele.ceriotti@epfl.ch}
\affiliation{Laboratory of Computational Science and Modeling, Institut des Mat\'eriaux, \'Ecole Polytechnique F\'ed\'erale de Lausanne, 1015 Lausanne, Switzerland}

\date{\today}%

\begin{abstract}

In the last few years several ``universal'' interatomic potentials have appeared, using machine-learning approaches to predict energy and forces of atomic configurations with arbitrary composition and structure, with an accuracy often comparable with that of the electronic-structure calculations they are trained on. 
Here we demonstrate that these generally-applicable models can also be built to predict explicitly the electronic structure of materials and molecules.
We focus on the electronic density of states (DOS), and develop PET-MAD-DOS, a rotationally unconstrained transformer model built on the Point Edge Transformer (PET) architecture, and trained on the Massive Atomic Diversity (MAD) dataset.
We demonstrate our model's predictive abilities on samples from diverse external datasets, showing also that the DOS can be further manipulated to obtain accurate bandgap predictions. 
A fast evaluation of the DOS is especially useful in combination with molecular simulations probing matter in finite-temperature thermodynamic conditions. 
To assess the accuracy of PET-MAD-DOS in this context, we evaluate the ensemble-averaged DOS and the electronic heat capacity of three technologically relevant systems: lithium thiophosphate (LPS), gallium arsenide (GaAs), and a high entropy alloy (HEA).
By comparing with bespoke models, trained exclusively on system-specific datasets, we show that our universal model achieves semi-quantitative agreement for all these tasks.
Furthermore, we demonstrate that fine-tuning can be performed using a small fraction of the bespoke data,  yielding models that are comparable to, and sometimes better than, fully-trained bespoke models. 

\end{abstract}

\maketitle

\newlength{\halfpage}
\setlength{\halfpage}{\columnwidth}

\section{Introduction}\label{sec:introduction}
\newcommand{\red}[1]{\textcolor{red}{#1}}
\newcommand{\black}[1]{\textcolor{black}{#1}}
\setlist[description]{font=\normalfont\bfseries}

Machine learning (ML) methods are rapidly gaining popularity in the field of computational materials science due to their ability to predict material properties at a fraction of the cost of traditional \textit{ab-initio} methods, while maintaining comparable levels of accuracy \cite{schmidt_recent_2019, schleder_dft_2019, wei_machine_2019}. ML models typically scale linearly with the system size, in contrast to \textit{ab initio} methods that are usually more costly and exhibit poorer scaling behaviour \cite{chandrasekaran_solving_2019}, which limits their usability for large or complex systems.

Early efforts in this domain were focused on highly specialized models, designed for specific properties in narrow regions of the chemical space. Examples of such early developments include interatomic potentials (IPs) \cite{unke+21cr, gigli2022thermodynamics} as well as models designed to predict bandgaps \cite{how_significance_2021, g_prediction_2022, zhuo_predicting_2018, how_dimensionality_2022}, charge densities \cite{lewis_learning_2021}, Hamiltonians \cite{nigam_equivariant_2022, li_deep-learning_2021}, nuclear magnetic resonance (NMR) spectra \cite{paruzzo_chemical_2018, C9SC03854J} or electronic density of states (DOS) \cite{ben_mahmoud_learning_2020, bang_accelerated_2021}. In recent years, there has been a shift towards developing universal models, i.e. models that are capable of generalizing well across a large fraction of the periodic table, spanning both molecules and extended materials \cite{Neumann2024, Batatia2023MACE-MP-0, Yang2024Mattersim}. However, these efforts have been largely focused on constructing universal ML interatomic potentials (MLIPs) to enable stable molecular dynamics simulations across diverse chemistries. Lately, there has been growing interest in building universal ML models to predict other material properties beyond energies and forces, such as bandgaps \cite{ruff_connectivity_2023, omee_scalable_2021, choudhary_atomistic_2021, chen+19cm}, Hamiltonians \cite{zhong_universal_2024, wang_universal_2024}, and the density of states \cite{kong_density_2022, fung_physically_2022, lee_predicting_2023}. 

Recently, a new universal MLIP, PET-MAD \cite{mazitov_pet-mad_2025}, has been introduced, reaching similar accuracies as existing state-of-the-art MLIPs for inorganic bulk systems while remaining reliable for molecules, organic materials and surfaces. The PET-MAD model employs the \textit{Point Edge Transformer} (PET) architecture \cite{pozd-ceri23nips}, a transformer-based graph neural network that does not enforce rotational symmetry constraints, but learns to be equivariant to a high level of accuracy through data augmentation. PET-MAD was trained on the small (containing fewer than 100,000 structures) but extremely diverse \textit{Massive Atomic Diversity} (MAD) dataset \cite{mazitov_massive_2025}. It encompasses both organic and inorganic systems, ranging from discrete molecules to bulk crystals. The dataset also includes randomized and heavily distorted structures to increase stability when performing complex atomistic simulations. Inspired by the success of the highly expressive PET architecture and highly diverse MAD dataset, we decided to apply this same combination to the prediction of the electronic density of states (DOS), a useful quantity for understanding the electronic properties of materials. 

The DOS quantifies the distribution of available electronic states at each energy level and underlies many useful optoelectronic properties of a material, such as its conductivity, bandgap and optical absorption spectra \cite{toriyama_how_2022, ashcroft_solid_1976}. These properties are highly relevant for applications like semiconductors and photovoltaic devices. Hence, the ability to easily obtain the DOS of a material can be instrumental for material discovery, paving the way for the development of better semiconductors or more efficient photovoltaics \cite{kong_density_2022}. Furthermore, the DOS can also enhance MLIPs by accounting for finite temperature effects, such as the temperature dependent electronic free energy \cite{chiheb2022} or electronic heat capacity \cite{Natasha2021}, thereby broadening their utility.

In this work, we present PET-MAD-DOS, a universal machine-learning model for predicting the DOS, based on the PET architecture and MAD dataset. \black{Uncertainty quantification (UQ) was also performed based on existing UQ methods \cite{bigi+24mlst, kell-ceri24mlst} to provide a measure for the accuracy of the DOS predictions at different energies.} We evaluate the performance of PET-MAD-DOS on atomistic benchmarks and ensemble quantities for a diverse set of scientifically interesting material systems, namely gallium arsenide (GaAs), lithium thiophosphate (LiPS), and high entropy alloys (HEA). We compare the ensemble quantities obtained using PET-MAD-DOS against bespoke models, i.e. PET models trained solely on those materials, and fine-tuned versions of PET-MAD-DOS for each material class. \black{These bespoke models have roughly half the test-set error of PET-MAD-DOS. The fact that a model specialized for a single material is only twice as accurate as our universal predictor is a testament to the robustness of PET-MAD-DOS. At the same time, having access to more accurate bespoke models trained on an entirely different specialized dataset allows us to assess the reliability of PET-MAD-DOS when using it in more complicated simulation workflows, whose validation by explicit electronic structure calculations would be prohibitively expensive.}

\section{Results}\label{sec:results}

This section covers the performance of PET-MAD-DOS, our foundation DOS model based on the PET architecture and trained on the MAD dataset. We report the details of the model and its training in the Methods (\autoref{sec:Methods}). We first showcase the performance and generalizability of PET-MAD-DOS by evaluating the DOS predictions on different subsets of the MAD dataset and several public datasets. 
Afterwards, we show that the predicted DOS can be used to obtain accurate predictions of the bandgap. Finally, we demonstrate the utility of our model on three case-study materials by evaluating ensemble quantities derived from MD trajectories. For these, we compared the performance of PET-MAD-DOS against that of (1) PET models trained solely on those systems and (2) the corresponding fine-tuned PET-MAD-DOS models.

\begin{figure*}
    \includegraphics[width=0.95\textwidth]{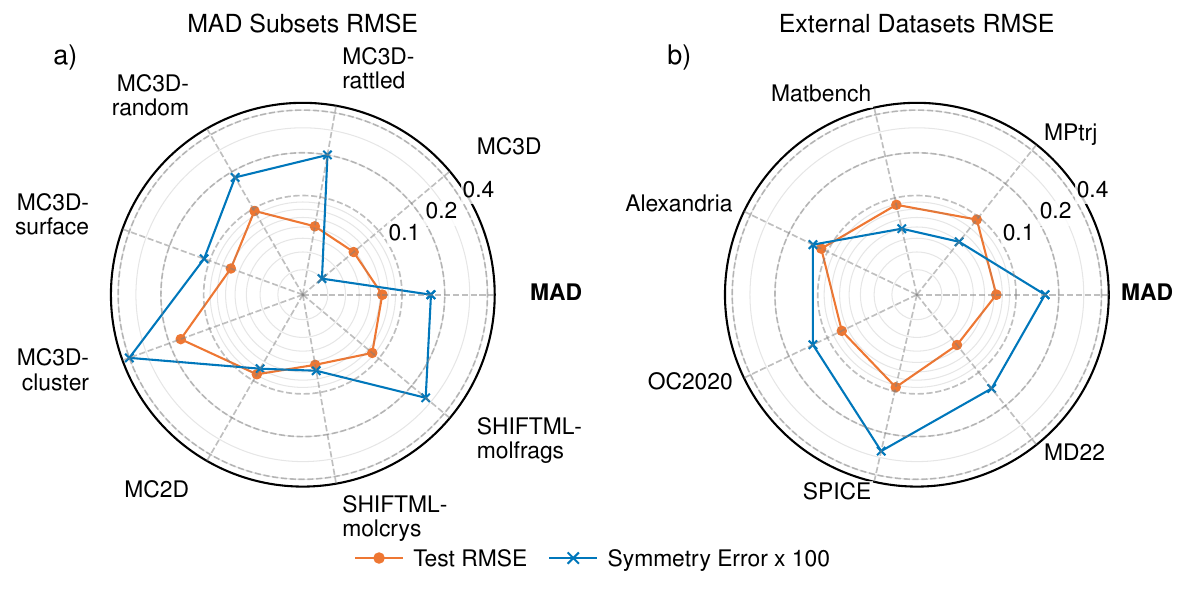}
    \caption{\label{fig:mad_benchmark}
    Root mean square error (RMSE) of the DOS predictions (orange-line) on the test set of the MAD dataset across the different subsets (a) and the external datasets (b). The blue line shows the rotational discrepancy, arising from the fact that PET is rotationally unconstrained. The symmetry error is multiplied by 100 to plot it on the same scale as the test RMSE, which is two orders of magnitude higher. Both the RMSE and the symmetry error are scaled based on the number of electrons in the system and have units of $\mathrm{eV^{-0.5}electrons^{-1}state}$.}
\end{figure*}

\begin{figure*}[t]
    \includegraphics[width=0.95\textwidth]{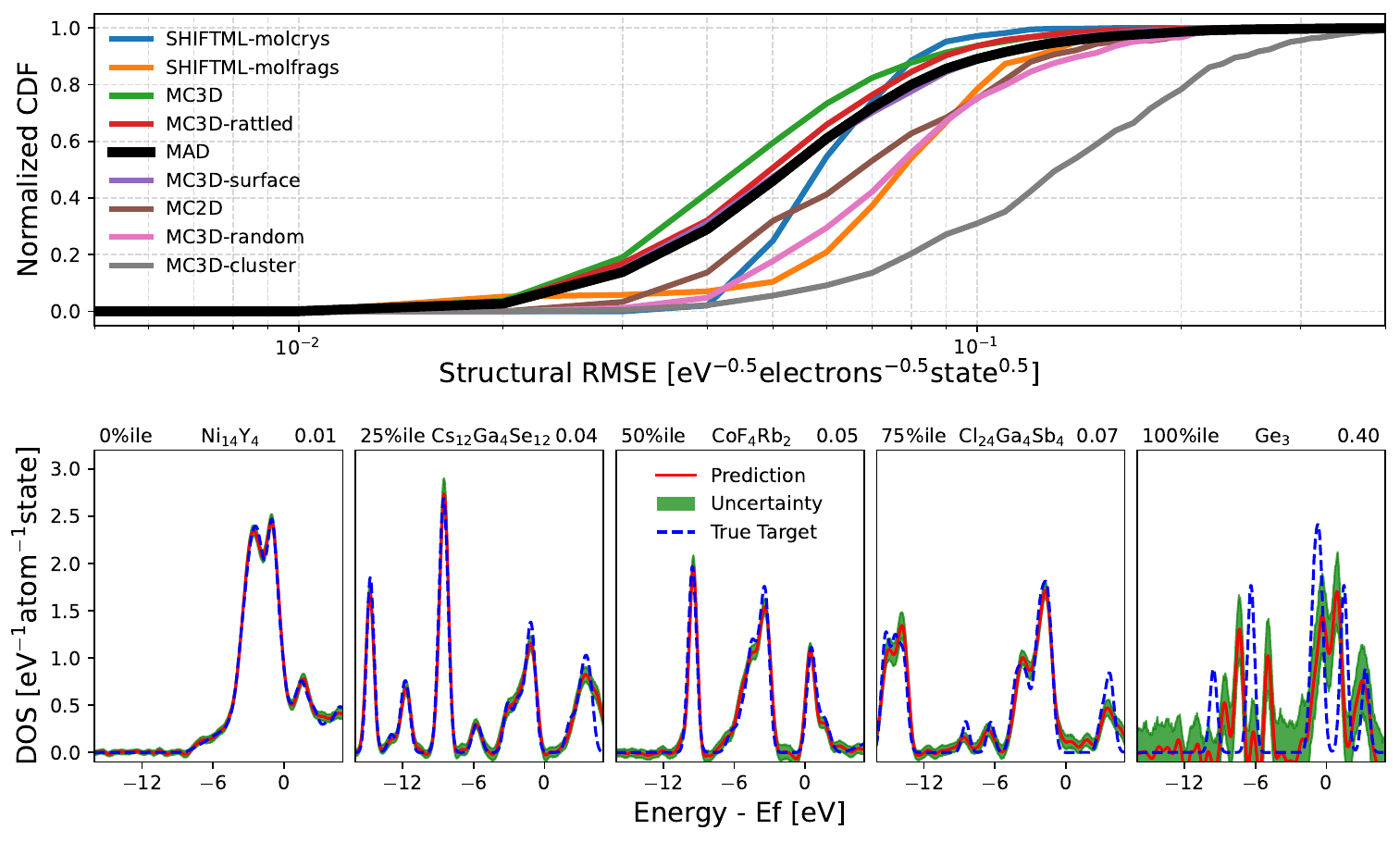}
    \caption{\label{fig:mad_errordistributions}
    Error distributions in the MAD test set. The top panel shows the normalized cumulative distribution function (CDF) of the RMSEs of each structure in each subset, represented by different colors, and the CDF of the entire MAD test subset in black. The bottom panel shows selected true DOS (blue dashed) /predicted DOS (red solid) comparisons from different parts of the MAD error distribution, for visual reference. The green areas represent the uncertainty associated with the DOS prediction as predicted by the calibrated last-layer prediction rigidity (LLPR) ensembles. The routine estimates the standard deviation $\sigma$ associated with the prediction at each energy channel. The range of the x axis has been truncated to ease visualization of the DOS predictions and its corresponding uncertainties. The RMSE corresponding to each subplot in the bottom panel is at the top right corner.}
\end{figure*}

\subsection{Model Performance}
\label{sec:performance}

We evaluate the performance of PET-MAD-DOS both on the MAD test set and on samples from other popular atomistic datasets, covering a broad spectrum of systems from bulk inorganic systems to drug molecules. The MAD dataset was originally developed as a compact dataset to train universal MLIPs, and is described in detail in Ref. \cite{mazitov_massive_2025}. It is divided into eight distinct subsets, which we summarize here:
\begin{description}
    \item[\textbf{MC3D} \& \textbf{MC2D}] Materials Cloud 3D (33596 structures) and 2D (2676 structures) crystal database respectively \cite{Huber2022MC3D, Campi2023MC2D}
    \item[\textbf{MC3D-rattled}] Structures generated by adding  Gaussian noise to the atomic positions of MC3D structures (30044 structures)
    \item[\textbf{MC3D-random}] Structures formed by randomizing the elemental composition of a subset of MC3D structures (2800 structures)
    \item[\textbf{MC3D-surface}] Surfaces obtained by cleaving a subset of MC3D structures cleaved along random crystallographic planes with low Miller index. (5589 structures)
    \item[\textbf{MC3D-cluster}] Clusters formed by randomly subselecting two to eight atoms from some MC3D structures. (9071 structures)
    \item[\textbf{SHIFTML-molcrys} \& \textbf{SHIFTML-molfrags}] Molecular crystals (8578 structures) and neutral molecular fragments (3241 structures) respectively from the SHIFTML dataset that is sampled from the Cambridge Structural Database \cite{Cordova2022, groom_cambridge_2016}
\end{description}

The samples from external datasets are recomputed using the MAD DFT settings to maintain consistency between training and evaluation data. 
They come from six sources: 

\begin{description}
    \item[\textbf{MPtrj}] Relaxation trajectories of bulk inorganic crystals dataset \cite{Deng2023}
    \item[\textbf{Matbench}] Bulk inorganic crystals from the Materials Project Database \cite{dunn_benchmarking_2020}
    \item[\textbf{Alexandria}] Relaxation trajectories of bulk inorganic crystals as well as 2D and 1D systems \cite{Schmidt2023}
    \item[\textbf{SPICE}] Drug-like molecules and peptides \citenum{east+23sd}
    \item[\textbf{MD22}] Molecular dynamics trajectories of peptides, DNA molecules, carbohydrates and fatty acids \cite{Chmiela2023}
    \item[\textbf{OC2020 (S2EF)}] Molecular relaxation trajectories on catalytically active surfaces \cite{Chanussot2021}
\end{description}

The errors of PET-MAD-DOS in these datasets are shown in \autoref{fig:mad_benchmark}, with further details of the error distributions in MAD illustrated in \autoref{fig:mad_errordistributions} which also provides a few representative example of DOS predictions, helping to relate the integrated errors to the visual quality of the predictions. Overall, the general performance trends of PET-MAD-DOS across the different datasets are similar to those of PET-MAD. For the MAD subsets, both models perform worst on the MC3D-random and MC3D-cluster subsets, likely due to the high chemical diversity in the subsets and the presence of several extreme cases of far-from-equilibrium configurations. The accuracy is especially poor for clusters, which have sharply-peaked DOS and often a highly nontrivial electronic structure. As shown in Figure \ref{fig:mad_errordistributions}, the error-distribution has a long tail, with a few high-error structures, but most of the structures having errors below 0.07 $\mathrm{eV^{-0.5}electrons^{-1}state}$. Considering the external datasets, \autoref{fig:mad_benchmark}b shows that PET-MAD-DOS performs best on MD22 and SPICE, which is consistent with the fact that the model performs better on the molecular part of the MAD dataset (SHIFTML subsets). Additionally, the performance of PET-MAD-DOS on the MAD dataset is comparable to that of the external datasets, highlighting both the chemical diversity of MAD and the ability of PET-MAD-DOS to capture the structure-property relationship in the extrapolative regime. Since the PET architecture does not impose any rotational constraints on the predictions, a rotated structure will not necessarily give the same prediction as the original structure despite the physical DOS being invariant to rotations. However, \autoref{fig:mad_benchmark} shows that the rotational discrepancy is two orders of magnitude smaller than the RMSE of the DOS. \black{Furthermore, recent works have shown that rotational discrepancies from rotationally unconstrained models have neglible impact on a model's performance in practical applications \cite{Langer_2024}.} Therefore, the lack of rotational constraint for PET-MAD-DOS does not impact the reliability of the model.

\black{In \autoref{fig:mad_errordistributions}, we also provide the uncertainties that have been quantified at each energy channel using the standard deviation of the calibrated last-layer prediction rigidity (LLPR) ensembles \cite{bigi+24mlst}. Information regarding the construction of the LLPR ensemble can be found in \autoref{sec:uq-details} of the Methods. The quantified uncertainties correspond well with the error made by the model for the structures shown on the bottom of \autoref{fig:mad_errordistributions}. Our LLPR-based uncertainty quantification (UQ) module is crucial for ensuring reliability in the model predictions, which is especially relevant for general-purpose models like PET-MAD-DOS as they are utilized in the ``edge'' cases where performance may deteriorate without warning. In particular for the DOS, the model's performance is inconsistent across energy channels, and thus our UQ module can be useful for identifying the model's confidence across different energy regions of the prediction.}

\subsection{Predicting the bandgaps}\label{sec:bandgap}

The bandgap plays a fundamental role in the optical and electronic properties of a material. Its magnitude provides insight into the electrical conductivity at different temperatures, as well as the wavelength of light that the material can absorb. Hence, predicting the bandgap can be very useful for material design in applications such as electronics, catalysis and photonics. 

\begin{figure}[t]
    \centering
    \includegraphics[width=0.45\textwidth]{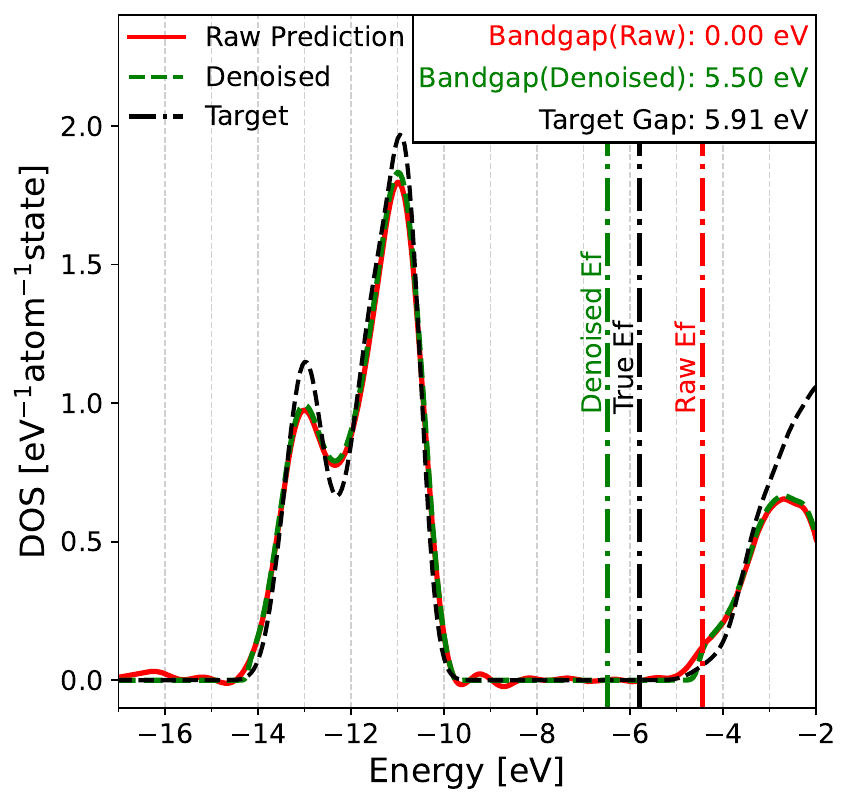}
    \caption{\label{fig:bandgap_cnn}
    Evaluation of the bandgap in MgCl$_2$, an insulator in the test set of MAD. The raw prediction of PET-MAD-DOS (solid red) is compared against that of the denoised prediction (dashed green) and true DOS (dash-dotted black). The colored vertical lines represent the Fermi level determined via integration of the corresponding DOS spectra. The target gap of 5.91eV represents the HOMO-LUMO gap obtained from the underlying DFT calculation while the other bandgaps are obtained from the corresponding DOS spectra, using a threshold of 0.1eV$^{-1}$atom$^{-1}$state.
    }
\end{figure}

In this work, we define the bandgap as the difference between the valence band maximum (VBM) and the conduction band minimum (CBM). To determine the bandgap from the DOS, one would normally first determine the Fermi level by finding the energy where the integrated DOS equals the total number of electrons in the system. The positions of the VBM and CBM can then be estimated to determine the bandgap. However, the application of this method to predicted DOS spectra poses several challenges. Although the DOS inside the bandgap should be zero, due to the use of Gaussian smearing to construct the target DOS, along with small prediction errors from the model, the DOS within the bandgap is often a small non-zero value. This introduces ambiguity in the choice of a threshold below which the DOS should be treated as zero. Another challenge is the determination of the Fermi level, which depends on the integrated DOS and therefore is very sensitive to accumulated errors. 
All these challenges are illustrated in \autoref{fig:bandgap_cnn} for MgCl$_2$, an insulator in the test set of MAD. The calculated Fermi level on the raw predicted DOS (red lines) is offset to the right of the gap by around 0.5 eV due to a slight underestimation of the integrated DOS. Since the Fermi level falls into a region with non-zero DOS, the physical interpretation is that MgCl$_2$ is a metal with no bandgap, which is qualitatively wrong. Even if the Fermi level was correctly determined, the oscillations in the predicted DOS (the most prominent one around -9 eV) would complicate the assessment of the gap's magnitude.

\begin{figure}[t]
    \centering
    \includegraphics[width=0.45\textwidth]{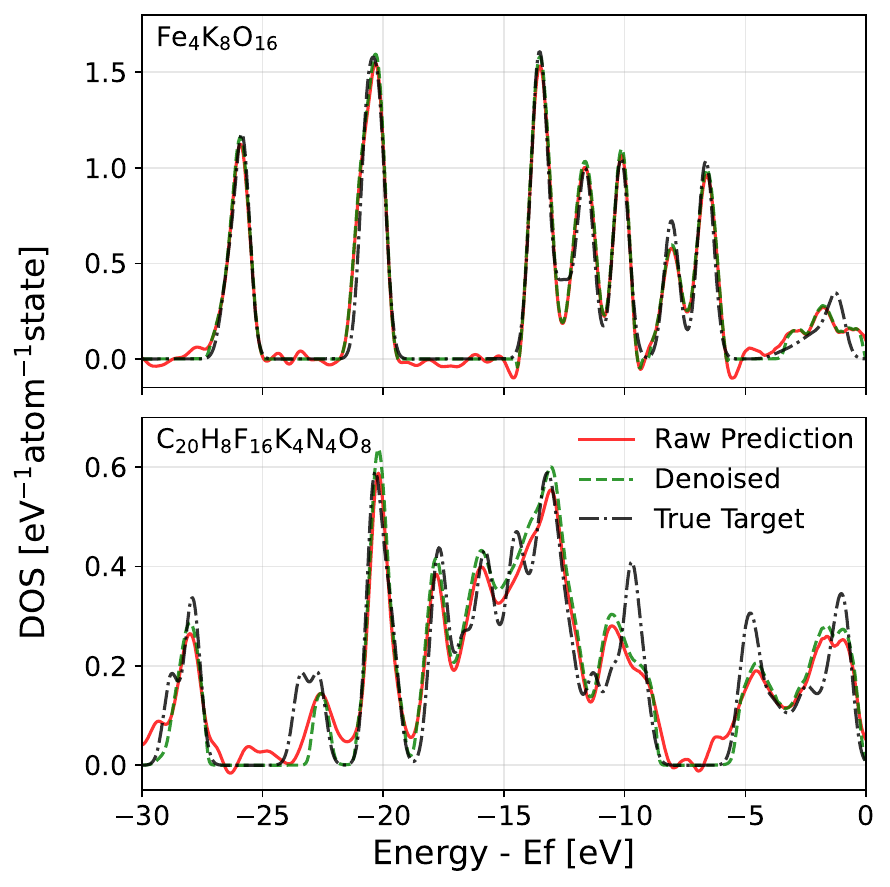}
    \caption{\label{fig:denoising}
    Demonstration of the effects of denoising on two sample predictions on the MAD test set. The raw prediction of PET-MAD-DOS (solid red) is compared against that of the denoised prediction (dashed green) and true DOS (dash-dotted black). The x-axis is truncated to enhance visualization of the differences between each DOS.
    }
\end{figure}

\begin{figure*}
    \includegraphics[width=\textwidth]{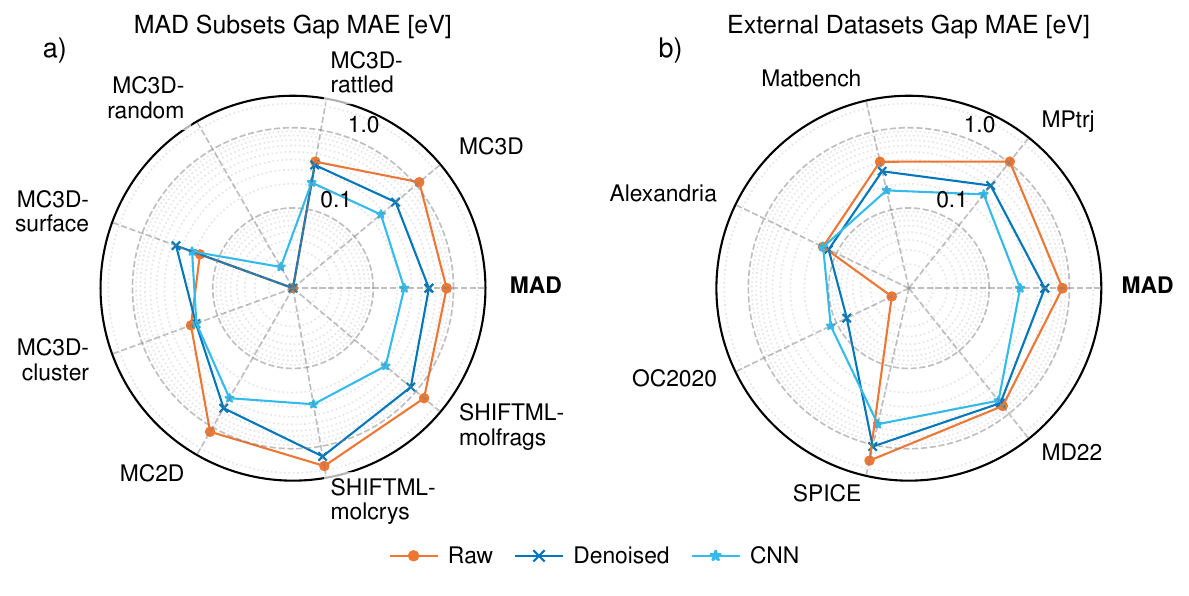}
    \caption{\label{fig:bandgaps}
    Comparison of the mean absolute error (MAE) of the bandgap predictions determined by physically interpreting the raw DOS prediction (orange), physically interpreting the denoised DOS prediction (dark blue), and applying a CNN on the raw DOS (light blue). The results are reported on the test set of the MAD dataset across the different subsets (a) and the external datasets (b). The MAE have units of eV. The values plotted in this figure are listed in Table I and Table II in the Supplementary Information.}
\end{figure*}

Given these issues, one may wonder whether the predicted DOS can be used to achieve the goals that motivated us to develop a DOS model in the first place. \black{To this end, we developed two solutions. The first solution involves passing the raw DOS prediction through a denoising filter to eliminate model noise in gap regions. The denoised DOS is also scaled such that the DOS integrates to the correct number of electrons at the Fermi level, which is predicted by a convolutional neural network (CNN) (See \autoref{sec: denoising} in the Methods for details). A demonstration of the denoising algorithm can be seen in both \autoref{fig:bandgap_cnn} and \autoref{fig:denoising}. Both figures show that the denoised prediction (green dashed line) exhibits virtually no oscillations in the gap regions, unlike the raw prediction (red solid line). For the case of MgCl$_2$ in \autoref{fig:bandgap_cnn}, the bandgap obtained from the denoised DOS is much better thanks to the improved Fermi level determination and higher quality DOS predictions in the gap. The second solution relies on a fully data-driven approach: the raw predicted DOS is passed through a CNN to predict the bandgap directly. The idea behind this solution is that a trained CNN should be able to find a way of dealing with noise that outperforms our handcrafted denoising algorithm, at the cost of being less elegant. For both approaches, the point that the CNN is applied is crucial. PET-MAD-DOS predicts atomic contributions that are summed over the atom indices to produce the total DOS. It is at this point where the CNN, which introduces non-linearities, should be applied. Applying it at the level of individual atomic environments would amount to making the assumption that a global quantity such as the bandgap and position of the Fermi level can be written as a sum of atomic contributions. For the same reasons, the denoising filter is applied to the total DOS and not to individual atomic contributions.}

\black{The performance of each method's bandgap predictions is displayed in \autoref{fig:bandgaps}, accompanied by tables I and II in the Supplementary Information. For the MAD test set, the CNN method achieves MAE errors that are roughly 4x lower than the raw predictions and 2x lower than the denoised predictions. In general, using the CNN method achieves better accuracies for secondary quantities. For instance, when estimating the DOS at the Fermi level, the MAEs of the raw predictions, denoised DOS and CNN method are 0.15, 0.13 and 0.10 $\mathrm{eV^{-1}atom^{-1}state}$ respectively. The results suggest that the CNN method yields superior performance, although the denoised DOS offers a reasonable alternative while keeping the workflow physically sound. Physical interpretability can be an advantage since it allows the derivation of additional properties from the same DOS without having to train more models. For example, we use the denoised DOS in \autoref{sec:electronic_heat_capacity} to compute the electronic heat capacity.}

The bandgap performance on the different MAD subsets and the external samples can also be seen in \autoref{fig:bandgaps}. The performance on bandgap does not necessarily follow the same trend as that of the DOS. Amongst the MAD subsets, the bandgap performance is best on the MC3D-random subset, where PET-MAD-DOS struggles to get good DOS predictions. A similar observation can be made for the Alexandria external dataset. On the other hand, the bandgap performance is poor on the SPICE and MD22 datasets, where PET-MAD-DOS performs well. This can be attributed to the distribution of bandgaps in those subsets. \black{For instance, the MC3D-random test subset consists entirely of conductors with no bandgap, and are thus easier to predict especially when using the raw or denoised DOS which tend to underestimate the bandgap.} A similar argument can be made for Alexandria, SPICE and MD22, where the error in each task correlates with its mean and standard deviation. In most cases the bandgap is predicted with an error around 100meV, which is comparable to the Gaussian smoothing we apply to construct the DOS, and smaller than typical DFT errors.

As a point of reference for the bandgap performance, we refer to the Matbench mp\_gap leaderboards, as of December 2025. Based on the CNN approach, \black{with a MAE and RMSE of 0.1900 and 0.3875 eV, it would be ranked 5th and 1st respectively.} However, we emphasize that this is only to give a point of reference regarding the performance of the model, and not to make a direct comparison with the models on the Matbench leaderboards. Firstly, the models on the Matbench leaderboards are trained on the Matbench dataset while our model is trained on the MAD dataset. Secondly, our evaluation is only done on a small sample of 140 structures, recomputed with MAD DFT settings while the Matbench leaderboard is based on the entire test subset, which we cannot use directly because it is computed with incompatible DFT settings. 

\subsection{Application to finite-temperature material simulations}
In addition to benchmarking PET-MAD-DOS on atomistic datasets, we demonstrate it in realistic applications by using it out-of-the-box as a general purpose model or as a foundation model to be fine-tuned. Towards that end, we used PET-MAD-DOS to predict the finite-temperature thermal-averaged DOS of two technologically relevant systems, namely Gallium Arsenide (GaAs) and Lithium thiophosphates (LPS), and to predict the electronic heat capacity of a high entropy alloy (HEA). Specific details with regards to the material simulations can be found in Section III of the Supplementary Information.

GaAs is a semiconductor with excellent physical and optoelectronic properties, making it well suited for photovoltaic devices for a wide range of applications \cite{sharma_gallium_2023}. The Ga-As system forms a simple binary phase diagram with metallic and semiconducting liquid and solid phases, making it an interesting system to use as a benchmark.

LPS have garnered great interest in the scientific community for their potential as electrolytes for solid-state batteries \cite{Kwade2018}. \ce{Li3PS4}, one of the most popular LPS, has been extensively studied and modelled computationally \cite{forrester2022, Gigli2024}. \ce{Li3PS4} has three main polymorphs, $\alpha$, $\beta$, and $\gamma$. The system is most stable in the $\gamma$ polymorph at room temperature but it transforms into the metastable $\beta$ polymorph at 573K and then into the $\alpha$ polymorph at 746K \cite{homma_crystal_2011}.  Although the $\gamma$ polymorph is a poor ionic conductor at room temperature, the $\beta$ polymorph has high ionic conductivity for Li$^+$, making it a promising candidate for a solid electrolyte. 

HEAs refer to systems composed of 5 or more metals in approximately equimolar proportions. These materials typically have desirable mechanical and catalytic properties \cite{yeh+04aem, cant+04msea, Sun2021, Katiyar2021}. However, building machine learning models to study HEAs and explore their composition space is often challenging due to the inherently high chemical diversity in these systems. They are often used in high-temperature applications, where thermal electronic excitations become relevant.

For all systems, we built a bespoke model, i.e. a PET model that is trained solely on the GaAs dataset from Imbalzano and Ceriotti \cite{imba-ceri21prm}, or the LPS dataset from Gigli \textit{et. al.} \cite{Gigli2024}, or a subset of the HEA25S dataset from Mazitov \textit{et. al.} \cite{mazitov_surface_2024}. All the datasets are recomputed with MAD DFT settings. Additionally, we also built a set of fine-tuned models by using the low-rank adaptive (LoRA) technique on the PET-MAD-DOS model on those datasets. Details on the fine-tuning procedure are discussed in \autoref{sec:fine-tuning-details}. The bespoke and fine-tuned models have typically half the test-set errors, and serve as an assessment of the accuracy of the zero-shot PET-MAD-DOS in these complex simulations that would be prohibitively expensive with DFT. 

\subsubsection{Test Set Performance}

To evaluate the performance of PET-MAD-DOS, the bespoke model and the LoRA fine-tuned model, we compare their accuracy on the test subset of those datasets in  \autoref{tab:rmse_comparison}. PET-MAD-DOS performs reasonably well out-of-the-box, achieving errors that are comparable with those computed on the MAD subsets. \black{The first thing to note is that PET-MAD-DOS errors are roughly twice as high as the errors of bespoke models in these systems. This is a common fact observed in other foundation models like MACE\cite{Batatia2023MACE-MP-0} and PET-MAD \cite{mazitov_pet-mad_2025} and does not diminish the utility of PET-MAD-DOS as a fast and inexpensive tool for qualitative DOS predictions for material systems across the periodic table.}

\begin{table}
\renewcommand{\arraystretch}{1.2}
\centering
\begin{tabular}{lccc}
\hline
\multicolumn{4}{c}{RMSE on Test subset [$\mathrm{eV^{-0.5}electrons^{-1}state}$]} \\
\hline
Material & PET-MAD-DOS & Bespoke Model & LoRA Model \\
\hline
GaAs     & 0.036 & \textbf{0.016} & 0.018 \\
LPS     & 0.064 & \textbf{0.027} & 0.030 \\
HEA   & 0.056 & 0.032 & \textbf{0.029} \\
\hline
\end{tabular}
\caption{Comparison of Test root mean squared error (RMSE) performance for bespoke, low rank adaptation (LoRA), and PET-MAD-DOS models on different systems. The best performing model for each material in indicated in bold.}
\label{tab:rmse_comparison}
\end{table}
\begin{figure*} [!ht]
    \includegraphics[width=\textwidth]{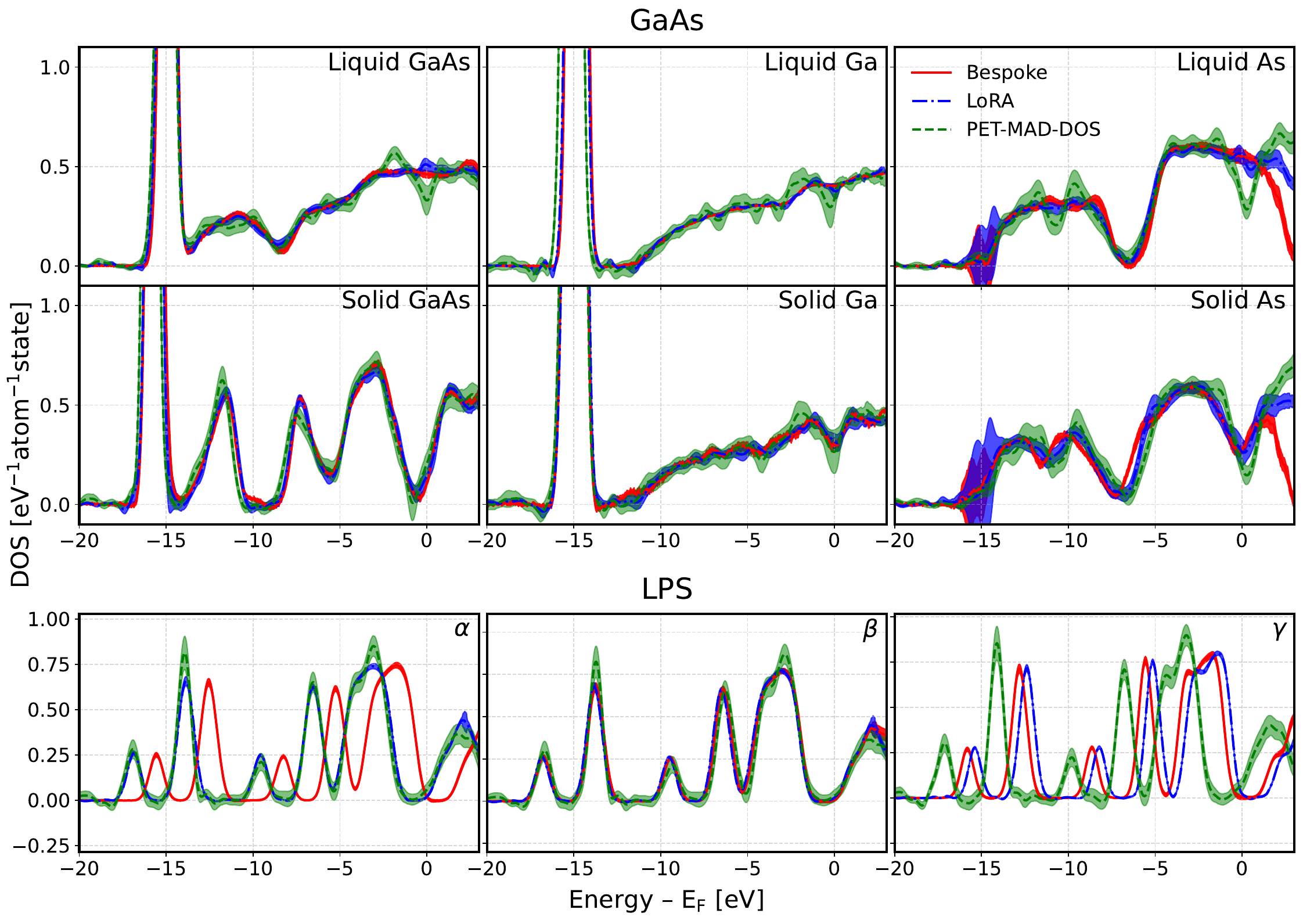}
    \caption{\label{fig:MD_predictions}
    Thermal-average DOS predictions of the MD trajectories of GaAs (top 2 rows) and LPS (bottom row) at different phases. The red solid lines represent the prediction of the bespoke model, the blue dash-dotted lines represent the prediction of the low rank adaptation (LoRA) model, and the green dotted line represents the prediction of PET-MAD-DOS. The colored areas represent the uncertainty associated with the DOS predictions of the corresponding model, obtained by propagating the uncertainties from each individual snapshot in the MD trajectory. In this procedure, the thermal-average DOS is computed for each member in the calibrated last-layer prediction rigidity (LLPR) ensemble, and the standard deviation across the ensemble members is taken as the uncertainty. Each system's phase is labelled at the top right corner of each subplot. The MD trajectories are obtained using a bespoke PET-MAD model. The energy axis shared between all systems is truncated to focus on the model's performance near the Fermi level, hiding the core and high energy states. A plot of the model predictions that includes the core states can be seen in Fig 5 of the Supplementary Information. For all subplots, the DOS is normalized with respect to the number of atoms in the system and the energy reference is set to the Fermi level determined based on each respective DOS prediction.} 
\end{figure*}

\black{Once PET-MAD-DOS is finetuned, it offers a performance similar or even better than that of the bespoke models.} The fine-tuned models are able to achieve bespoke accuracies without significant impact to their performance on the MAD dataset (Table III of the Supplementary Information). Furthermore, based on the learning curves in section IV of the Supplementary Information, the fine-tuned models have good performance even in the low-data regime, where they clearly outperform the bespoke ones. For the LPS and HEA datasets, the fine-tuned models are able to achieve bespoke accuracies using only 20\% of the training data.

\subsubsection{Thermal-Average DOS}

In addition to evaluating the models on their test set performance, we also compare each model's ability to compute the thermal-average DOS along molecular dynamics (MD) trajectories of GaAs and LPS in different phases.
Studying phase transitions or interfaces requires atomistic models of thousands or more atoms, for which computing thermal-averages of the DOS is beyond the capabilities of conventional electronic structure methods. Deringer \textit{et. al.} \cite{deringer_origins_2021} have previously combined MLIPs with ML models for the DOS to reveal electronic properties in large amorphous silicon systems up to 100k atoms, proving the potential of the approach to reach unprecedented system sizes. However, their study relied on bespoke models. In this section, \black{we demonstrate that similar results can also be obtained using only universal models, eliminating the need to train bespoke models, which can be computationally expensive during both the training and data generation phase.}

For GaAs, we used NVT MD trajectories of Ga, GaAs, and As in both solid and liquid phases generated with the bespoke interatomic potential in Ref. \cite{mazitov_pet-mad_2025}. For the solid systems, the MD simulations were performed at 150K, 750K and 550K for Ga, GaAs, and As respectively. Meanwhile, for the liquid systems, the temperatures are 450K, 2250K, and 1650K for the Ga, GaAs, and As systems. For both solids and liquids, the temperatures are chosen to be well into the solid or liquid phases, so as to avoid spurious phase transitions due to the limitations of the reference DFT energetics. The simulations were performed for 4ns, using a timestep of 4fs. 

For LPS, we used the MD trajectory generated by the bespoke interatomic potential in Ref. \cite{mazitov_pet-mad_2025}. The trajectories for the three LPS phases were performed in the NpT ensemble at 400K for a quasi-cubic cell containing 768 atoms at a pressure of zero bar. The trajectories were run for 3 ns, sampled every 20 fs.

\autoref{fig:MD_predictions} shows that PET-MAD-DOS is generally able to qualitatively predict the same DOS profile as the bespoke model, up to roughly 3eV above the Fermi level. \black{The LLPR module acts as a good estimate of the model confidence, as evidenced by the good overlap between the uncertainties of all three models. In this case, the profiles are a thermal average of model predictions across a MD trajectory, so we need to propagate uncertainty. To do so, we first compute the thermal-average predicted by each LLPR ensemble member. We then take the mean over LLPR ensemble members to get the final prediction, and use the standard deviation as a measure of uncertainty.} It is crucial to note that the decay of the DOS above the Fermi level for the bespoke model is likely not physical as it arises due to the limited number of eigenstates in the DFT calculations used for the training set. For LPS, the predictions are observed to be offset relative to one another when aligned at the Fermi level. This is attributed to the difficulty in determining the Fermi level for a predicted DOS spectra as highlighted in \autoref{sec:bandgap}.
However, the shape of the DOS profiles still closely matches that of the LoRA and bespoke models. \black{Along with the overlapping uncertainties, this highlights the fact that PET-MAD-DOS is able to yield good qualitative results out of the box in practical applications.}

\label{sec:electronic_heat_capacity}
\subsubsection{Electronic Heat Capacity}

\begin{figure}[t]
    \centering
    \includegraphics[width=0.5\textwidth]{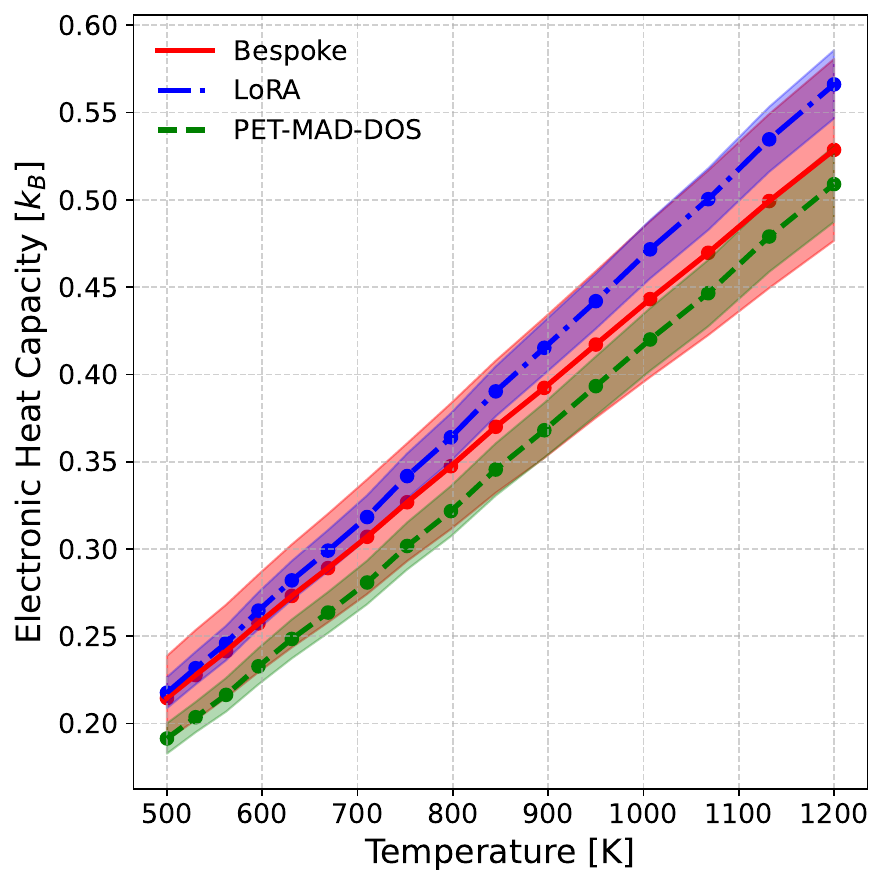}
    \caption{\label{fig:HEA_MD}
    Constant pressure electronic heat capacity derived from the thermal-average DOS of the HEA system at 16 different temperatures from 500K to 1200K. The red solid line represent the prediction of the bespoke model, the blue dash-dotted line represent the prediction of the low rank adaptation (LoRA) model, and the green dotted line represents the prediction of PET-MAD-DOS. The colored areas represent the uncertainty associated with the DOS predictions of the corresponding model, obtained by propagating the uncertainties from each individual snapshot in the MD trajectory. In this procedure, the heat capacity is computed for the denoised prediction of each member in the calibrated last-layer prediction rigidity (LLPR) ensemble, and the standard deviation across the ensemble members is taken as the uncertainty.} 
\end{figure}

For HEAs, we evaluate the quality of the thermal-averaged DOS by using it to obtain the electronic heat capacity of a prototypical CoCrFeMnNi alloy. The electronic heat capacity can be particularly relevant at high temperatures, making it important for HEAs used in high temperature applications. 

In this work, we calculated the electronic heat capacity from the HEA MD trajectories obtained using PET-MAD in Ref. \cite{mazitov_pet-mad_2025}. The trajectories were obtained using a combination of replica-exchange molecular dynamics run with Monte-Carlo atom swaps. The simulation was performed with 16 replicas for 200 ps in the NPT ensemble using a 2 fs timestep at zero pressure and using a logarithmic temperatures grid ranging from 500K to 1200K.

\black{To derive the heat capacity from the DOS, we used the denoised DOS as described in \autoref{sec:bandgap} instead of the raw DOS predictions, due to its higher physical interpretability. First, the thermal-averaged DOS was computed as the average of the denoised predictions along the MD trajectory.} Then, the electronic contribution to the internal energy, $U^\mathrm{el}$, was computed under the rigid band approximation as highlighted in Ref. \cite{lopa+21prm}. The electronic heat capacity was then calculated as the derivative of $U^\mathrm{el}$ with respect to temperature using a finite difference scheme. Further details on the computation of the electronic heat capacity can be found in Section III of the Supplementary Information. \black{The uncertainties for the heat capacities are propagated by computing the heat capacity for each member in the calibrated LLPR ensemble, taking the mean as the predicted heat capacity and the standard deviation as the uncertainty.} The results are shown in \autoref{fig:HEA_MD}, where it can be observed once again that PET-MAD-DOS performs well, being able to capture semi quantitatively the trend between heat capacity and temperature. \black{Furthermore, the overlapping uncertainties reflect good agreement between all 3 models.}

\section{Discussion}
PET-MAD-DOS consistently achieves semiquantitative predictions of the DOS and properties that can be extracted from it. Despite being trained on a small dataset and having a moderate number of parameters, it performs well across a broad spectrum of material classes, even on structures from external datasets. \black{The generalizability of PET-MAD-DOS exceeds that of other universal DOS models \cite{kong_density_2022, fung_physically_2022} which are trained on datasets consisting solely of inorganic systems.} Furthermore, its performance out-of-the-box is only a factor of two worse than that of bespoke models trained on a medium-sized dataset for a specific class of material. \black{This allows PET-MAD-DOS to yield results that are close to those of the bespoke models even in practical applications, highlighting the efficacy of PET-MAD-DOS as a general purpose tool for DOS predictions. Furthermore, with the uncertainty quantification module based on an LLPR ensemble, it is also possible to have a reliable estimate of the model's error for both the DOS and the derived quantities at a relatively low cost. If the projected error is unsatisfactory,} the performance can also be enhanced for particular applications by using PET-MAD-DOS as a foundation model to be fine-tuned for enhanced accuracies. The performance of these fine-tuned models is close to the bespoke models, sometimes outperforming them on their own validation domain. Learning curves show that fine-tuning works well with only about 100 additional structures, \black{requiring far less data than bespoke models.} Furthermore, the fine-tuned model still retains stable predictions for the more general datasets.

Although the PET architecture employed does not enforce rotational constraints, PET-MAD-DOS is still able to predict the DOS with a high level of rotational invariance, with the rotational variability being 2 orders of magnitude smaller than the accuracy of the model. PET-MAD-DOS is built and integrated within the \texttt{metatensor} \cite{METATRAIN} ecosystem, allowing the model to be easily accessible and for the training procedure to be easily replicated. Based on the accessibility, versatility and utility of PET-MAD-DOS, we believe that it can serve as a useful tool for materials discovery, especially in applications that require explicit information on the electronic structure.

\section{Methods}\label{sec:Methods}
In this section, we introduce details with regard to dataset construction, model architecture, loss functions for training and model evaluation, model training procedure, bandgap model architecture, and model fine-tuning, and uncertainty quantification. Further details regarding the MD simulations and hyper-parameters of the model can be found in sections III and V of the Supplementary Information.

\subsection{Dataset construction}\label{sec:mad_details}

As the MAD dataset was primarily constructed to fit MLIPs, it was computed using a minimal number of energy bands. The energy range in which the DOS is well defined, based on the eigenvalues calculated, varies widely across the dataset. To increase data representation at energies above the Fermi level, a small subset of 850 structures was recalculated using four times the number of valence bands in the system. These structures are 750 monoelemental systems from the MC3D and MC3D-rattled subsets, together with the 100 structures that possess the lowest energy cutoff in the entire MAD dataset. Including these recomputed structures improves the DOS predictions in the high energy range, as displayed in Section VI of the Supplementary Information. Additionally, for bandgap benchmarking purposes, a small random subset comprised of 140 structures was taken from the Matbench dataset and recomputed with the same DFT settings outlined in Ref. \cite{mazitov_pet-mad_2025}. 

The calculations above were performed using the Quantum Espresso v7.2 package \cite{gian+09jpcm}, under a non-magnetic setting with the PBEsol exchange-correlation functional. The pseudopotentials used were obtained from the standard solid-state pseudopotentials library (SSSP) v1.2 (efficiency set) \cite{pran+18npjcm}, using the highest settings for the plane-wave and charge density cutoffs across all 85 elements present in the MAD dataset (110 Ry and 1320 Ry respectively). The Marzari-Vanderbilt-deVita-Payne cold smearing \cite{marz+99prl} was used, with a spread of 0.01 Ry. For structures with periodicity, a fine k-point spacing of 0.125 $\pi$ Å$^{-1}$ was used in every periodic dimension while only one k point was used for the non-periodic dimensions. See Ref. \cite{mazitov_massive_2025} for a detailed discussion of the makeup of the MAD dataset. 

The target DOS for a structure, $\mathrm{DOS}_A^{Q}(E)$, is then built via Gaussian smearing of the eigenvalues at each k-point and projecting it on a uniform energy grid as follows:
\begin{align}
     \label{DOS}\mathrm{DOS}_A^{Q}(E) &= \sum_{n \in \mathrm{bands}}\sum_\textbf{k}w_k \  g(E-\epsilon_n(\textbf{k})) \\
     g(x) &= \frac{1}{\sqrt{2\pi\sigma^2}}e^{-\frac{x^2}{2\sigma^2}},
\end{align}
where $N_A$ represents the number of atoms in the structure. $\epsilon_n(\textbf{k})$ represents the eigenvalues at each k point, with the energy reference set to the Fermi level determined by Quantum Espresso in the quantum chemical calculation. $w_k$ represents the weight of the k-point in the Brillouin zone integral. $\sigma$ is a Gaussian smearing parameter which is set to 0.3eV, determined by comparing the constructed DOS of a sample structure against that of the same structure computed with a finer k-grid. $E$ is the energy grid, which is a uniform grid \black{containing 4806 points} from -149.65eV to 80.65eV, representing 1.5eV below and above the lowest and highest eigenvalue cutoff in the original MAD dataset (excluding recalculated structures). The lowest eigenvalue cutoff is the lowest eigenvalue in the dataset while the highest eigenvalue cutoff is the minimum energy of the highest energy band in the dataset.

\subsection{PET model}\label{sec:pet_details}
The {Point Edge Transformer} (PET) \cite{pozd-ceri23nips} architecture combines both transformers and graph neural networks by using transformers in the message-passing layer. For every system, a directed graph is built by defining atoms as nodes and directed edges connect atoms within a specified cutoff radius. Feature vectors $f_{ij}^l$ are then built on each directed edge between atoms $i$ and $j$. These feature vectors serve as the messages that will be passed in the message-passing layer, $l$. The dimensionality of $f_{ij}^l$ is fixed and is defined by a hyperparameter of the architecture, $d_{\mathrm{PET}}$. In each message-passing layer, a transformer is used to perform a permutation-covariant sequence-to-sequence transformation. The transformer takes as input all feature vectors $f_{ij}^l$, for a given central atom $i$ and layer $l$, and outputs the corresponding feature vectors $\{ f_{ij}^{l+1} \}_j$  for the next layer $l + 1$. This step also incorporates structural and chemical information regarding the central atom, such as the 3D positions of the neighbors and chemical species. After going through all the message-passing layers, all feature vectors $f_{ij}^l$ are then used as inputs for a final feed-forward network. \black{The output of the final feed-forward network} is summed across bonds $ij$ and layers $l$ and represents the final target property, \black{an array with size 4806 depicting the DOS in this case.} To obtain better expressivity, the PET architecture does not impose any rotational constraints, allowing a single layer to theoretically access virtually unlimited body orders and angular resolution. To address the lack of rotational symmetry constraints, data augmentation is employed for the model to learn the rotational behaviour of the target, i.e. invariant for the case of the DOS. 

\black{In this work, the only change made to the original PET architecture is at the last layer of the final feed-forward network, which is modified to give 4806 outputs, representing the size of the DOS array, instead of 1.} For a more detailed description of the architecture and specific operations, the reader can refer to the original PET publication \cite{pozd-ceri23nips}. 

\subsection{Training and evaluation functions}\label{sec:loss functions}

A simple mean squared error loss function is unable to properly reflect the underlying physical constraints of the DOS as a machine learning target, especially in a highly chemically diverse dataset where each calculation has a different energy cutoff in the eigenvalues. To account for the lack of an absolute energy reference in bulk systems \cite{kleinman_comment_1981}, we use a loss function that is agnostic to the energy reference of the prediction and the target. For this, we compute the loss only on the energy reference that minimizes the prediction error. We define the self-aligning loss, $AL$, for a single structure $A$ as such:
\begin{align}
    \label{MSE}\textrm{MSE}(y(E), \hat{y}(E)) &= \int_{E_{min}}^{E_{max}} dE\ (y(E) - \hat{y}(E))^2 \nonumber \\  & \quad+ \int_{G_{min}}^{E_{min}} dE\ y(E)^2 
\end{align}

\begin{align}
    \label{ALoss} AL_A(\mathbf{W}) = \min_{\Delta \in {0,1,\ldots, \chi}}\bigg[\mathrm{MSE}\bigg(\textrm{DOS}^\mathbf{W}_A(E + (\Delta \times e)),\nonumber \\ \textrm{DOS}^Q_A(E)\bigg)\bigg].
\end{align}
$E_{min}$ and $E_{max}$ denote the energy minimum and maximum of the evaluation window. $\textrm{DOS}^Q_A(E)$ represents the true DOS for structure $A$ while $\textrm{DOS}^\mathbf{W}_A(E)$ represents the predicted DOS for structure $A$ given model parameters $\mathbf{W}$. $\chi$ is an integer that denotes the maximum number of grid points the energy reference can shift by and $e$ represents the energy grid interval. $G_{min}$ refers to the minimum energy of the prediction grid and the second term in the Eq. \eqref{MSE} essentially fits the DOS predictions below $E_{min}$ to zero to reflect that there are no states below the minimum eigenvalue. This arises due to the fact that this minimization procedure requires the model to predict the DOS in a wider energy grid, resulting in $G_{min} \leq E_{min}$. The optimization algorithm then searches for the continuous subset within the prediction, corresponding to the size of the target, that minimizes the MSE. Based on preliminary testing, we have set $\chi$ to 200, corresponding to the prediction grid being 10eV wider. This is similar to the adaptive energy reference used in Ref. \citenum{how_adaptive_2025}, with the exception that the loss is now fully minimized at every epoch instead of being optimized simultaneously with the model weights, but the energy reference can only shift in integer multiples of the energy grid interval. By restricting the search space to only integer multiples, it circumvents the need to compute derivatives or build splines of the DOS during the minimization procedure. Additionally, we were able to exploit full vectorization to evaluate the loss for all values of $\Delta$ simultaneously, ensuring that the minimization procedure obtains the global minima. 

Although every system, in principle, has an infinite number of eigenvalues at every k-point, electronic structure calculations consider only a finite number of them. Due to this restriction, calculating the DOS based on the method outlined in \autoref{sec:mad_details} will result in a sharp unphysical drop in the DOS to zero, past the maximum computed eigenvalue. This impacts the reliability of the DOS targets computed near the highest computed eigenvalue. To account for this during model evaluation and training, we set $E_{max}$ in \eqref{ALoss} for each structure to 0.9 eV, corresponding to 3$\times$ the smearing value, below the minimum energy of the highest energy band across every k-point. Since MAD was computed with a minimal number of energy bands, a large number of structures have a low $E_{max}$, with some $E_{max}$ values being lower than the Fermi level. Hence, it is not feasible to simply set the $E_{max}$ of all structures to the minimum $E_{max}$ in the dataset. Additionally, due to the wide range of $E_{max}$ in the dataset, there is an uneven distribution of data across the energy grid. This results in highly oscillatory predictions at higher energy levels due to insufficent data in those regions. These oscillations can contaminate predictions during deployment if the structure contains atomic environments that comes from two training structures with very different $E_{max}$ (Section VI of Supplementary Information). To address these oscillations, we introduce a gradient loss, $GL$, that imposes a mean squared penalty on the gradient of the predictions, determined via finite differences, outside $E_{max}$. The gradient loss for a single structure, $A$, is:
\begin{align}
    GL_A(\mathbf{W}) = \int_{E_{max}}^{G_{max}} dE \ \bigg(\frac{d\,\mathrm{DOS}_A^{\mathbf{W}}\big(E + (\Delta_{opt} \times e)\big)}{dE} \bigg)^2,
\end{align}
where $G_{max}$ represents the maximum energy of the prediction grid and $\Delta_{opt}$ is the optimal shift determined via (\ref{ALoss}).

In addition, we also include the loss on the cumulative DOS, $CL$, similar to Ref. \cite{fung_physically_2022, benm+20prb}. The loss on the cumulative DOS for a single structure, $A$, is expressed as:
\begin{align}
    CL_A(\mathbf{W}) =  \int_{E_{min}}^{E_{max}} dE \ \bigg(\textrm{cDOS}^\mathbf{W}_A\big(E + (\Delta_{opt} \times e)\big) \nonumber \\-\  \textrm{cDOS}^Q_A(E)\bigg)^2
\end{align}
where cDOS represents the cumulative DOS function. The final loss that the model is trained on is as follows:
\begin{align}
    \label{final loss}L(\mathbf{W}) = \frac{1}{N}\sum_A \frac{1}{N_A}\bigg(AL_A(\mathbf{W}) + \alpha GL_A(\mathbf{W}) \\ + \beta CL_A(\mathbf{W})\bigg),
\end{align}
where $N$ refers to the number of structures in the training set and $N_A$ denotes the number of atoms in structure A. The loss is normalized with respect to the number of atoms in each structure to make the loss independent of structure size. $\alpha$ and $\beta$ are hyperparameters used to scale $GL$ and $CL$ respectively. In this work, $\alpha$ and $\beta$ are set to $10^{-4}$ and 2 based on preliminary tests. 

For evaluation, the RMSE is also normalized to account for the difference in the number of electrons represented by the DOS in the dataset:
\begin{align}
    n_A &= \int_{E_{min}}^{E_{max}} dE \ \mathrm{DOS}^Q_A(E)\\
    \label{RMSE}RMSE &= \sqrt{\frac{1}{N}\sum_A \frac{AL_A(\mathbf{W})}{n_A}}
\end{align}
where $N$ represents the number of structures in the evaluation set. $n_A$ represents the number of electrons represented in the target DOS.

We evaluate the symmetry error as the standard deviation of the DOS predictions of 38 rotated copies of the each structure, based on a Lebedev angular grid with a degree of 8. The standard deviations are only computed up to the point where the DOS target is defined so that it can be compared to the RMSE of the DOS predictions. The formula for the symmetry error, $\sigma_A^{rot}$, is as follows:

\begin{align}
    \sigma_A^{rot} &= \sqrt{\frac{1}{38}\sum_{i=1}^{38} \frac{1}{n_A}\int_{E_{min}}^{E_{max}} dE \ (DOS_A^i (E) - DOS_A^\mu(E) )^2},
\end{align}
where $i$ represents the index of the rotated copies, $A$ represents the structure, $DOS_A^i$ represents the prediction on the $i$th rotated copy of structure $A$ and $DOS_A^\mu(E)$ represents the mean prediction of structure $A$ across all rotations. The symmetry error is normalized by the number of electrons so that it can be meaningfully compared against the RMSE in \eqref{RMSE}.

\subsection{Training of PET-MAD-DOS}\label{sec:pet-mad_details}

Each one of the eight subsets in the MAD dataset were split into training, validation, and test sets in a 8:1:1 ratio. We perform a hyperparameter search over the five points on the Pareto-front of PET-MAD \cite{mazitov_pet-mad_2025} and select the hyperparameters that yield the best balance of performance and accuracy. The results are detailed in Section VII of the Supplementary Information, \black{where we also report the computational cost of PET-MAD-DOS.} The resulting optimal hyperparameters are the same as those in PET-MAD, with a cutoff radius of 4.5Å, 2 message-passing layers, each comprising of two transformer layers with a token size of 256 and 8 heads in the multi-head attention layer. The output multi-layer perceptron contains 512 neurons, which are fed to a linear layer to give 4806 outputs, corresponding to the DOS at each energy channel. This results in a total of 8,625,226 parameters in the model. Model training was performed using the PyTorch framework and the \texttt{metatrain} package \cite{METATRAIN} on 1 NVIDIA H100 GPU with a batch size of 16 structures for a total of 760 epochs, taking roughly 72 hours. For model training, the Adam \cite{Adam} optimizer was used, with an initial learning rate (LR) of $10^{-4}$, using a warmup of 100 epochs that increases the LR linearly from 0 to $10^{-4}$. Afterwards, a LR scheduler was employed to half the LR every 250 epochs. 

\subsection{CNN model Specifications}\label{sec:bandgapmodeldetails}

For the CNN models used to predict secondary quantites like the bandgap, Fermi level and DOS($\mathrm{E_F}$) model, we utilize a simple 1D convolutional neural network (CNN) for univariate sequential input. The model takes the \black{raw} PET-MAD-DOS prediction of a structure as input and is composed of four sequential convolutional blocks followed by two fully connected layers. Each convolutional block contains a convolutional layer with 64 output channels and a SiLU activation function, and a 1D max pooling layer with a kernel size of 4. The kernel size of the convolution layer in the first, second, third and fourth block is 32, 16, 8 and 8 respectively. The two fully connected layers contains 1024 neurons each, with the SiLU activation function to produce a scalar output representing either the target. The model is trained on the mean squared error (MSE) against the DFT targets, using the Adam optimizer with an initial LR of $10^{-4}$ and 100 warmup epochs that increases the LR linearly from 0 to $10^{-4}$. Early stopping is implemented to stop model training if the MSE on the validation set does not decrease after 50 epochs. The model is trained using the Pytorch framework on 1 NVIDIA H100 GPU with a batch size of 16 for roughly 150 epochs, taking around 30 minutes. During evaluation, the ReLU activation function is applied to the predictions of the bandgap model to remove unphysical negative bandgap values. 

\subsection{Prediction Denoising}\label{sec: denoising}

\color{black}As highlighted in \ref{sec:bandgap}, relying on a physical interpretation of the raw predicted DOS for the Fermi level and bandgap requires extremely high DOS accuracies and minimal noise in the gap. As this is difficult to achieve under the current training approach, an additional prediction denoising step was applied on the DOS predictions to obtain a DOS that can be physically interpreted.

Firstly, a CNN model was trained, as described in \ref{sec:bandgapmodeldetails}, to predict the position of the Fermi level of a structure based on the raw predicted DOS. Then, a 1-D Gaussian filter, with a standard deviation, $\sigma$, of 0.3 $eV$ was applied on the raw predicted DOS as follows,
\begin{align}
    \mathrm{DOS}_G(E) & =  \int_{G_{min}}^{G_{max}} \mathrm{DOS}_{\text{pred}}(\tau) G(E- \tau) d\tau \\
    G(E) & = \frac{1}{\sigma\sqrt{2\pi}} \exp\left(-\frac{E^2}{2\sigma^2}\right),
\end{align}
where $\mathrm{DOS}_G$ represents the filtered DOS and $\mathrm{DOS}_{\text{pred}}$ represents the raw DOS prediction. Next, the filtered DOS is passed through a modified sigmoid function,
\begin{align}
    f(x) =  \frac{1}{1+e^{-a(x-b)}},
\end{align}
where the additional constants $a$ and $b$ determine the inflection point and slope of the sigmoid function. In this work, we chose $a$ to be 0.1 and $b$ to be 100. The output of the modified sigmoid function, $\beta$, is then used as a multiplier on the DOS output to obtain a thresholded DOS.
\begin{align}
    {\mathrm{DOS_{thresh}}}(E) &= \mathrm{DOS}_{\mathrm{pred}}(E) * f(\mathrm{DOS}_G(E))
\end{align}
In the last step, the thresholded DOS is then scaled such that the physical Fermi level of the DOS lie on the same point as that predicted by the Fermi level CNN, described in the first step.
\begin{align}
    n &= \int_{G_{min}}^{\epsilon^{CNN}_F} {\mathrm{DOS_{thresh}}}(E)\\
    {\mathrm{DOS_{clean}}} &= \frac{n_{elec}}{n}{\mathrm{DOS_{thresh}}}(E)
\end{align}
where $\mathrm{DOS_{clean}}$ represents the final denoised DOS output, $n_{elec}$ refers to the number of electrons in the neutral system (excluding the ones in the pseudopotential), and $\epsilon^{CNN}_F$ refers to the Fermi level of the system predicted by the CNN model described in the first step.\color{black}

\subsection{Fine-tuning}\label{sec:fine-tuning-details}

The popular low-rank adaption (LoRA) method \cite{LoRA} was employed to fine-tune the pre-trained PET-MAD-DOS models for specific applications. LoRA was selected for its efficiency and ability to reduce the impacts of \textit{catastrophic forgetting}, which refers to a fine tuned model losing its predictive capabilities on its base dataset. Instead of fine tuning all the model weights as in conventional fine-tuning, LoRA instead trains an additional set of parameters while leaving the original model weights untouched. These parameters are comprised of two low-rank matrices which are added to each attention block of the model, scaled by a regularization factor that controls the influence of the matrices on the model's weights. Through tuning the rank of the matrices and the regularization factor, a model can be fine tuned to achieve better performance in specific applications without compromising the generalizability of the model. In this work, we use the same LoRA parameters as PET-MAD, namely a rank of 8 and the regularization factor set to 0.5.

LoRA-fine-tuned models retain varying degree of accuracy (see the Table III of the Supplementary Information for details) on the generic structures from the MAD dataset, while providing performance comparable to that of a bespoke model, even in the low data regime for certain systems. Hence, we recommend the use of LoRA when fine-tuning PET-MAD-DOS for a specific application.

\subsection{Uncertainty quantification}~\label{sec:uq-details}

\black{To perform uncertainty quantification (UQ) for the PET-MAD-DOS model, we employed the last-layer prediction rigidity (LLPR) method by Bigi et al.~\cite{bigi+24mlst}, which computes uncertainties as the inverse of the prediction rigidity.~\cite{chon+23jctc,chongPredictionRigiditiesDatadriven2025a} The fact that DOS is a vectorial prediction target presents limitations in the originally proposed UQ approach: the last-layer features of each structure used for DOS prediction is fixed for all energy channels, and calibration factors are obtained ``globally'' across the entire dataset, resulting in a fixed uncertainty profile for all structures, only scaled differently based on the relative magnitude of the prediction rigidity. We therefore initialize a last-layer ensemble of 128 models with the weights sampled following Eq.~25 of Ref.~\cite{bigi+24mlst}. We perform further calibration of the ensemble weights with a Gaussian negative log-likelihood loss as done in Kellner and Ceriotti~\cite{kell-ceri24mlst}, resulting in a UQ profile that is far more informative and accurate (see Figure 11 of the Supplementary Information). Furthermore, the UQ profile also accurately reflects the adaptive evaluation window used in the loss function for training. The model uncertainty increases significantly when extrapolating the DOS to high energies, as observed in Figure 12 of the Supplementary Information.}

\begin{acknowledgments}
The Authors would like to thank Davide Tisi for kindly providing the molecular dynamic trajectories for LPS. The authors would also like to thank Guillaume Fraux and Philip Loche for their contributions to the development of the \texttt{metatrain} infrastructure. They are also grateful to the current and past members of the Laboratory of Computational Science and Modeling who contributed to the software infrastructure that supported this work.
Computation for this work relied on resources from the EPFL HPC platform (SCITAS). 

\paragraph*{Funding:}

MC and WBH acknowledge the funding from the European Research Council (ERC) under the European Union’s Horizon 2020 research and innovation programme (grant agreement No 101001890-FIAMMA).
MC and PF acknowledge funding from the MARVEL National Centre of Competence in Research (NCCR), funded by the Swiss National Science Foundation (SNSF, grant number 182892)
AM and MC acknowledge support from an Industrial Grant from BASF.
SP and FB were supported by a project within the Platform for Advanced Scientific Computing (PASC).
MK, SC, and MC acknowledge support by the Swiss National Science Foundation (grant ID 200020\_214879).

\paragraph*{Author contributions:}

WBH worked on the development of the loss function and integration into \texttt{metatrain}, training PET-MAD-DOS, bespoke and LoRA models, developed and trained the CNN models and denoising workflows, performed the accuracy, performance, and speed benchmarks, calculated the ensemble-average DOS and electronic heat capacity on MD trajectories of the three material systems, DFT re-computations of a subset of MAD and DFT computations for the Matbench sample.
PF guided the methodology for the determination of the electronic heat capacity, denoising and bandgap from the DOS.
SC integrated the UQ strategy described in~\ref{sec:uq-details} into \texttt{metatrain}, performed the calibration of LLPR-based last-layer ensembles for UQ, and aided the evaluation and analysis of the prediction uncertainties of PET-MAD-DOS.
AM worked on the creation of the MAD dataset, sample selection and DFT calculation of the external datasets, implemented the LoRA training procedure on \texttt{metatrain}, integrated PET-MAD-DOS within the PET-MAD repository and performed the MD simulations of surface segregation in CoCrFeMnNi.
FB worked on implementation of the \texttt{metatrain} infrastructure for training and evaluating the PET-MAD-DOS model, and supported the development of the infrastructure for UQ. 
MK performed on the MD simulations of Ga, GaAs, and As in the solid and liquid phases. 
SP developed the original version of PET architecture and the shift invariant loss function.
MC designed and guided the project, and provided theoretical support.
All authors contributed to the writing of the manuscript.
\paragraph*{Competing interests:}
There are no competing interests to declare.

\paragraph*{Data and materials availability:}

The MAD dataset, benchmarks, and simulation input files are available as a record \cite{mazitov_2025_xdsbt-a3r17} on the Materials Cloud Archive \cite{tarl+20sd}. 
A dataset containing the curated DOS data, including also the structures recomputed with a larger number of empty states and training scripts for PET-MAD-DOS, uncertainty quantification, and finetuning will be made available upon publication.
The pre-trained PET-MAD-DOS model, along with the necessary dependencies, is available on the PET-MAD repository at \url{https://github.com/lab-cosmo/pet-mad}.

\end{acknowledgments}

\clearpage

\renewcommand{\thefigure}{S\arabic{figure}}
\setcounter{figure}{0}
\renewcommand{\thesection}{S\arabic{section}}
\setcounter{section}{0}

\onecolumngrid
\part*{Supplementary Information}
\onecolumngrid

\section{Details of benchmarking subsets selection}

The performance of PET-MAD-DOS was evaluated on samples from several popular atomistic datasets computed with MAD DFT settings as reported in subsection 2.1 of the main text. In this section, we detail the method in which the samples were obtained from the respective datasets. 

\begin{description}
    \item[$\mathbf{MPtrj}$] MACE-MP-0 validation subset, reduced to 153 structures after removing four 1D wire structures 
    \item[$\mathbf{Matbench}$] 140 randomly sampled structures from the Matbench mp\_gap dataset 
    \item[$\mathbf{Alexandria}$] 200 randomly sampled structures, 50 from Alexandria-2D, 50 from Alexandria-3D-gopt, and 100 from the Alexandria-3D subset.
    \item[$\mathbf{SPICE}$] 100 randomly sampled neutral molecules from the SPICE dataset.
    \item [$\mathbf{MD22}$] 149 structures, obtained by randomly sampling 25 structures from each of the seven subsets of the MD22 dataset (Ac-Ala3-NHMe, AT-AT, DHA, Stachyose, AT-AT-CG-CG, Buckyball-Catcher, double-walled-nanotube), and then cleaned of non-converged cases.
    \item [$\mathbf{OC2020}$] 89 structures obtained by sampling 100 structures from the OC2020-S2EF training dataset and then cleaned of non-converged cases
\end{description}

Wherever applicable, structures containing elements that are not contained in the MAD dataset are excluded from the random selection. Aside from the Matbench sample, the remaining samples are obtained from Ref. \cite{mazitov_pet-mad_2025}. All samples are computed using MAD DFT settings outlined in subsection 4.1 of the main text and Ref. \cite{mazitov_massive_2025}.

\clearpage

\section{Comparison of bandgap determination methods}

As mentioned in the main text, it is difficult to obtain reliable bandgap estimates from the 
DOS, especially if it is constructed using Gaussian smearing. This can be attributed to the fact that the DOS is not exactly zero but a small value in the gap, which raises ambiguity regarding the threshold at which the DOS should be treated as zero. Due to the small DOS value in the gap, small errors in the DOS can significantly affect bandgap predictions. To tackle this issue, we propose two solutions. One solution involves passing the raw DOS output of PET-MAD-DOS through a machine-learned denoising approach outlined in Section 2.2 and 4.6 of the main text. This approach significantly reduces the noise in the gap region and enhances the determination of the Fermi level, resulting in more reliable bandgap predictions from the DOS. Alternatively, we also propose the use of a simple CNN model to learn the bandgap from the raw output of PET-MAD-DOS to make the determination process more robust. In the tables below, we compare the performance of these methods in determining the bandgap of the system, as an additional point of comparison, we also report the results when trying to determine the bandgap from the true DOS using the same threshold. As a note, the error for the true DOS is not zero due to the fact that the true DOS is constructed using Gaussian smearing and the bandgap is defined as the HOMO-LUMO gap. With the exception of the CNN, the bandgap determination method uses a DOS threshold of $10^{-1}\mathrm{eV^{-1}atom^{-1}state}$, and lower values are considered as zero for the purposes of bandgap determination. Threshold values below $10^{-1}\mathrm{eV^{-1}atom^{-1}state}$ resulted in the raw DOS approach yielding no bandgaps for nearly every structure.

\begin{table}[h]
\renewcommand{\arraystretch}{1.2}
\centering
\begin{tabular}{lccccccccc}
\hline
\multicolumn{10}{c}{Bandgap Test MAE/RMSE on different subsets of MAD [eV]} \\
\hline
 & MAD-Test & MC3D & MC2D & Rattled & Random & Surface & Cluster & MolCrys & MolFrags \\
\hline
Raw DOS & 0.82 & 1.13 & 1.16 & 0.40 & \textbf{0.00} & \textbf{0.17} & 0.23 & 1.78 & 1.36 \\
Denoised & 0.49 & 0.47 & 0.53 & 0.36 & \textbf{0.00} & 0.36 & \textbf{0.19} & 1.34 & 0.82 \\
CNN & \textbf{0.24} & \textbf{0.27} & \textbf{0.38} & 0.22 & 0.02 & \textbf{0.22} & \textbf{0.19} & \textbf{0.29} & \textbf{0.32} \\
\hline
True DOS & 0.28 & 0.29 & 0.27 & 0.18 & 0.00 & 0.03 & 0.13 & 0.75 & 0.65 \\
\hline
\hline
Mean Gap & 1.08 & 1.33 & 1.29 & 0.40 & 0.00 & 0.10 & 0.21 & 2.88 & 3.54\\
\hline

\hline

\end{tabular}
\caption{\label{tab:bandgap_MAD_comparison}
    Bandgap MAE of the different bandgap determination methods on the MAD test subsets. The CNN approach uses a convolutional neural network to predict the bandgap of the system via the raw DOS output from PET-MAD-DOS. The other methods predicts the bandgap from a given DOS spectra via a physical interpretation, first determining the Fermi level via integration and determining the bandgap based on the DOS values around the Fermi level. For this, the DOS threshold was set to $10^{-1}\mathrm{eV^{-1}atom^{-1}state}$, below which the DOS was considered to be zero for the purposes of determining the bandgap. The boldface values refer to the approach that led to the best bandgap prediction using only the predicted DOS. In the last row, we report the mean bandgap across every structure in each subset.
    }
\end{table}

\begin{table}[h]
\renewcommand{\arraystretch}{1.2}
\centering
\begin{tabular}{lcccccc}
\hline
\multicolumn{7}{c}{Bandgap MAE on external benchmarks [eV]} \\
\hline
& MPtrj & Alexandria & SPICE & MD22 & OC2020 & Matbench \\
\hline
Raw DOS & 1.04 & 0.15 & 1.60 & 0.75 & \textbf{0.02} & 0.41 \\
Denoised & 0.43 & \textbf{0.13} & 1.06 & 0.68 & 0.07 & 0.31 \\
CNN & \textbf{0.31} & 0.15 & \textbf{0.55} & \textbf{0.62} & 0.12 & \textbf{0.18} \\
\hline
True DOS & 0.24 & 0.11 & 0.96 & 0.54 & 0.03 & 0.19 \\
\hline
\hline
Mean Gap & 0.71 & 0.15 & 3.2 & 3.2 & 0.02 & 0.88\\
\hline

\hline

\end{tabular}
\caption{\label{tab:bandgap_ext_comparison}
    Bandgap MAE of the different bandgap determination methods on samples of the external benchmarks. The CNN approach uses a convolutional neural network to predict the bandgap of the system via the raw DOS output from PET-MAD-DOS. The other methods predicts the bandgap from a given DOS spectra via a physical interpretation, first determining the Fermi level via integration and determining the bandgap based on the DOS values around the Fermi level. For this, the DOS threshold was set to $10^{-1}\mathrm{eV^{-1}atom^{-1}state}$, below which the DOS was considered to be zero for the purposes of determining the bandgap. The boldface values refer to the approach that led to the best bandgap prediction using only the predicted DOS.In the last row, we report the mean bandgap across every structure in each subset. }
\end{table}

From both \autoref{tab:bandgap_MAD_comparison} and \autoref{tab:bandgap_ext_comparison}, we can see that the CNN approach typically performs best, followed by using the denoised predictions. In the cases where the raw DOS performs extremely well, namely in the MC3D-Random subset of the MAD test set and OC2020, the reason is because these structures tend to be conductors with no bandgap, and the raw DOS tends to severely underestimate the bandgap. The converse is true when the mean bandgap is very high, like in SPICE and MD22, where the raw-DOS prediction performs very poorly. It is important to point out that due to the tendency to underestimate gaps, the bandgaps obtained by the raw DOS are all zeroes for the benchmark samples from OC2020 and even MD22, which generally has high bandgaps. This underscores the importance of postprocessing methods, like denoising the predictions or using a CNN.

\clearpage

\section{Simulations}

In this section, we provide further details regarding the parameters with which the finite temperature material simulations have been conducted. For these systems, molecular dynamics were performed using LAMMPS \cite{THOMPSON2022108171} with either the PET-MAD machine learning interatomic potential (MLIP) or the PET bespoke MLIP to obtain the relevant trajectories. The reference DFT level of PET-MAD and the bespoke machine learning potentials are PBEsol, consistent with the level of theory of the PET-MAD-DOS model.

\subsection{Gallium arsenide}
For the Gallium/Arsenide (Ga/As) material systems, we computed thermal averages of the GaAs DOS in the NVT ensemble, employing the bespoke MLIP in Ref. \cite{mazitov_pet-mad_2025} for the pure phases system (Ga, GaAs, and As) in both the solid and liquid states. The bespoke MLIP was trained on the same GaAs dataset as discussed in the main text, which samples across the binary phase diagram of GaAs, including surfaces and highly distorted structures \cite{ImbalzanoGaAs2021}. Further details regarding the model and dataset can be found in the original publications.  For the MD simulations, the liquid structures of Ga, GaAs, and As were generated using Packmol \cite{Packmol}. The solid Ga crystal structure was selected from the Materials Project database \cite{Jain2013}, while solid GaAs \cite{GaAs_solid}, and solid black As \cite{Smith01011975} were obtained from the Inorganic Crystal Structure Database \cite{Zagorac:in5024} - ICSD (As: ICSD-70100 , GaAs: ICSD-610540) (ICSD release 2025.1). For all systems, we relaxed the positions of the initial structures and performed MD simulations for 4 ns employing a 4fs timestep and a Nose-Hoover thermostat \cite{PhysRevA.31.1695}.

For Ga, the liquid system contains 384 atoms in a cell with size 18.12 \AA $\times$ 23.25 \AA $\times$ 18.37 \AA. The solid system contains 64 atoms in a cell of size 8.86 \AA $\times$ 15.20 \AA $\times$ 9.11 \AA. MD was performed on these systems at 450K and 150K for the liquid and solid systems respectively.

For GaAs, the liquid system is composed of 256 Ga and 256 As atoms, in a cubic cell with length 23.49 \AA, and MD was performed at 2250K. The solid system has 32 Ga and 32 As atoms in a cubic cell with length 11.31 \AA, and MD was performed at 750K.

For As, the liquid simulation was performed on a 19.14 \AA $\times$ 16.58 \AA $\times$ 21.23 \AA unit cell with 300 As atoms at 1650K. The solid simulation was performed on a 7.30 \AA $\times$ 8.93 \AA $\times$ 22.00 \AA unit cell with 64 As atoms at 550K.

All simulation temperatures were chosen well separated from the experimental melting points.
\subsection{Lithium thiophosphate}

For the LPS molecular dynamics simulations, we use the same trajectory as the one in the Ref. \cite{mazi+24jpm} generated using the bespoke LPS PET MLIP. The simulations were performed according to the protocol in the reference publication.

The LPS simulations were performed using a bespoke PET model in the NpT ensemble for a quasi-cubic 768-atom cell in the $\alpha$, $\beta$, and $\gamma$ phase, with a constant isotropic pressure of $p$ = 0. The MD trajectory used in this work was performed at 400K, for 3ns with a timestep of 2fs. Further details can be found in the reference publication \cite{Gigli2024}.

\subsection{High-entropy alloys}

For the HEA MD simulations, we also use the same trajectory as that in Ref. \cite{mazi+24jpm}. The simulations were performed according to the protocol outlined in the reference publication \cite{mazitov_surface_2024}.

The simulations were performed using the PET-MAD model on a CoCrFeMnNi alloy surface slab with a \textit{fcc} lattice in the (111) orientation and a $7\times 7\times 11$ supercell containing 539 atoms. Relaxation of both structure and composition of the surface was performed with replica-exchange molecular dynamics run with Monte-Carlo atom swaps with 16 replicas for 200 ps in the NPT ensemble using a 2 fs timestep at zero pressure and logarithmic temperature grid ranging from 500K to 1200K.

To compute the electronic heat capacity, we use an approach adapted from the work of Lopanitsyna \textit{et. al.} \cite{lopa+21prm}. The electronic contribution to the internal energy of the system is calculated from the DOS based on the following equation,

\begin{align}
    U_\textrm{DOS}^\textrm{el} &= \int_{-\infty}^{\infty} dE\ E\times \mathrm{DOS}^T(E) f(E-\mathrm{E}^T_\mathrm{F}, T) \nonumber \\& \quad - \int_{-\infty}^{\mathrm{E}^0_\mathrm{F}} dE\ E\times \mathrm{DOS}^T(E), 
\end{align}

where the $\mathrm{DOS}^T$ represents the thermal-average DOS over a particular temperature T. $f(E, T)$ represents the Fermi-Dirac distribution, and $\mathrm{E}^T_\mathrm{F}$ represents the Fermi level determined at a particular temperature T. The electronic heat capacity, $C_p$, is then computed as the derivative of $U_\textrm{DOS}^\textrm{el}$ with respect to temperature using a finite difference scheme with 2 points and a temperature interval of 1K.

\clearpage

\section{Learning curves}\label{sec:learning-curves}

\subsection{PET-MAD-DOS}

\begin{figure}[h]
    \centering
    \includegraphics[width=0.5\textwidth]{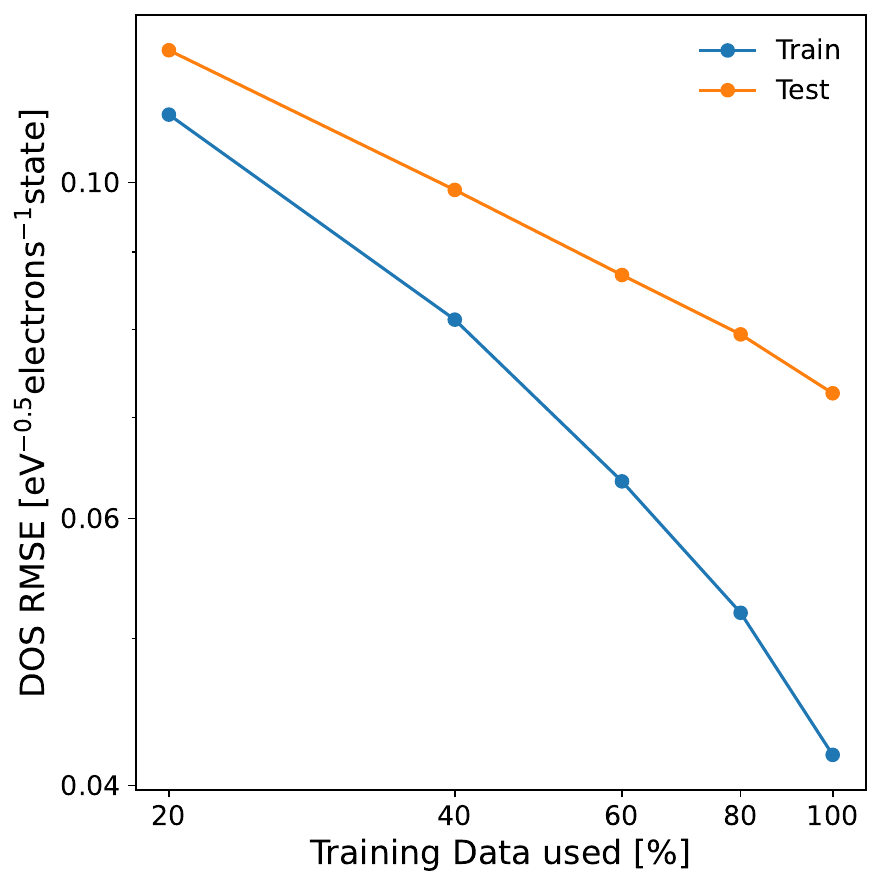}
    \caption{Learning curves of PET-MAD-DOS. The amount of training data, randomly sampled from the MAD training set, is represented on the x-axis as a percentage and the Test DOS RMSE is represented on the y-axis.}
    \label{fig:pet-mad-dos_lc}
\end{figure}

The learning curve of PET-MAD-DOS is shown in \autoref{fig:pet-mad-dos_lc}. Each model is trained on a subset of the MAD dataset, obtained by randomly selecting the corresponding percentage of training structures from each subset, and then combined and shuffled. From the figure, it can be observed that the model's test performance steadily improves with the size of the training set, and has yet to saturate. This indicates that the model's performance can be further enhanced by increasing the size of the training set. 

\subsection{Gallium arsenide}

\begin{figure}
    \centering
    \includegraphics[width=0.5\textwidth]{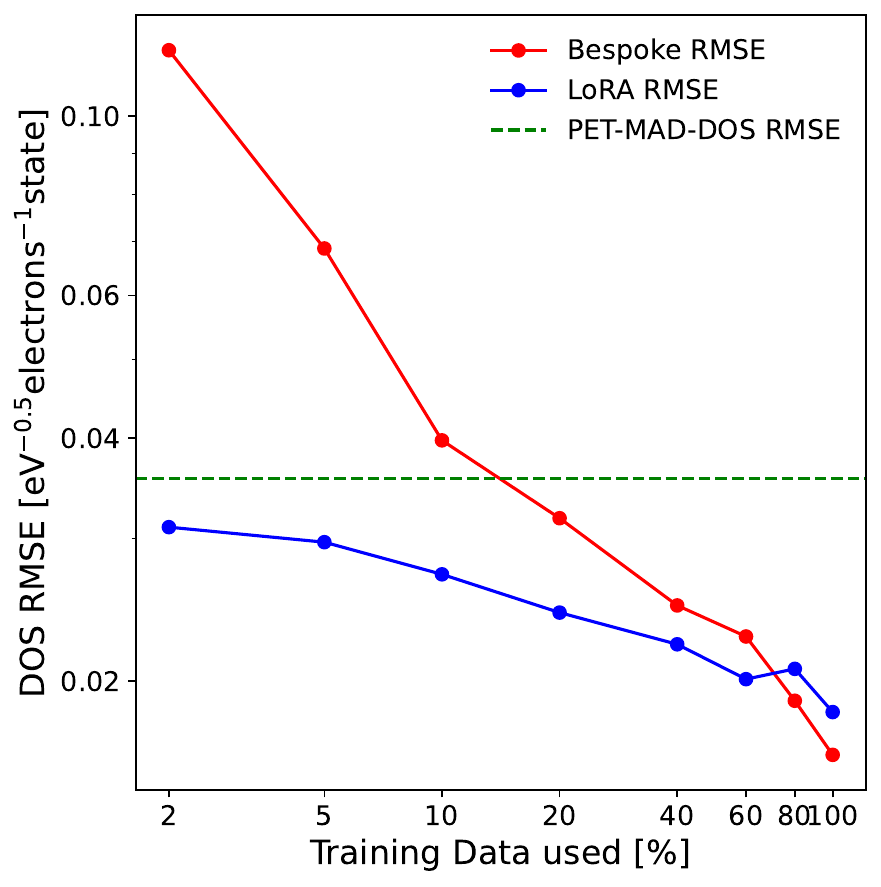}
    \caption{Learning curves for the GaAs dataset, comparing the performance of the bespoke model and the LoRA fine-tuned model and that of the PET-MAD-DOS model. The amount of training data, randomly sampled from the GaAs training set (1417 structures), is represented on the x-axis as a percentage and the Test DOS RMSE is represented on the y-axis.}
    \label{fig:GaAs_lc}
\end{figure}

The learning curves of the bespoke model and LoRA fine-tuned model for GaAs are shown in \autoref{fig:GaAs_lc}. From the figure, it can be seen that the test performance of both models has yet to saturate, and that the LoRA fine-tuned models tend to outperform bespoke models, especially in the low data regime. Furthermore, the bespoke models only outperform PET-MAD-DOS when the training set is at least 10\% (142 structures) of the dataset. 

\subsection{Lithium thiophosphate}

\begin{figure}
    \centering
    \includegraphics[width=0.5\textwidth]{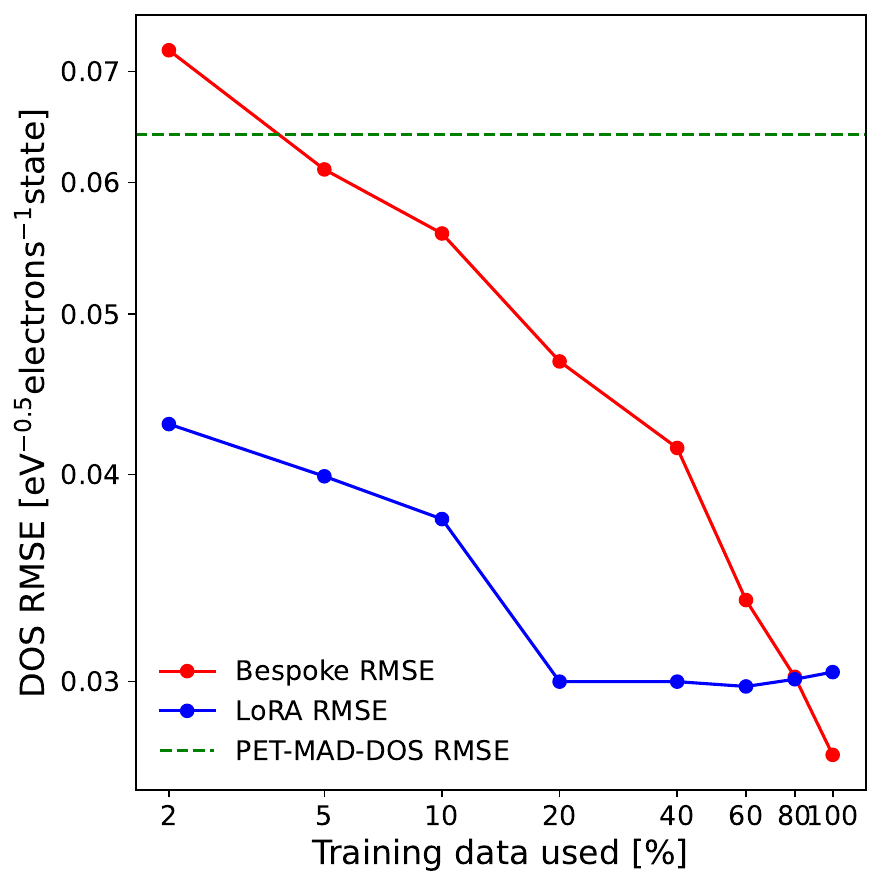}
    \caption{Learning curves for the LPS dataset, comparing the performance of the bespoke model and the LoRA fine-tuned model and that of the PET-MAD-DOS model. The amount of training data, randomly sampled from the LPS training set (1940 structures), is represented on the x-axis as a percentage and the Test DOS RMSE is represented on the y-axis.}
    \label{fig:LiPS_lc}
\end{figure}

\autoref{fig:LiPS_lc} shows the learning curves for the Li$_{3}$PS$_{4}$ (LPS) dataset. Interestingly, the test performance for the Lora- fine-tuned models has saturated at 20\% of the training data while the bespoke models have yet to saturate. This indicates that using LoRA finetuning on PET-MAD-DOS allows one to obtain performant models with a smaller dataset. 

\subsection{High-entropy alloys}

\begin{figure}[h]
    \centering
    \includegraphics[width=0.5\textwidth]{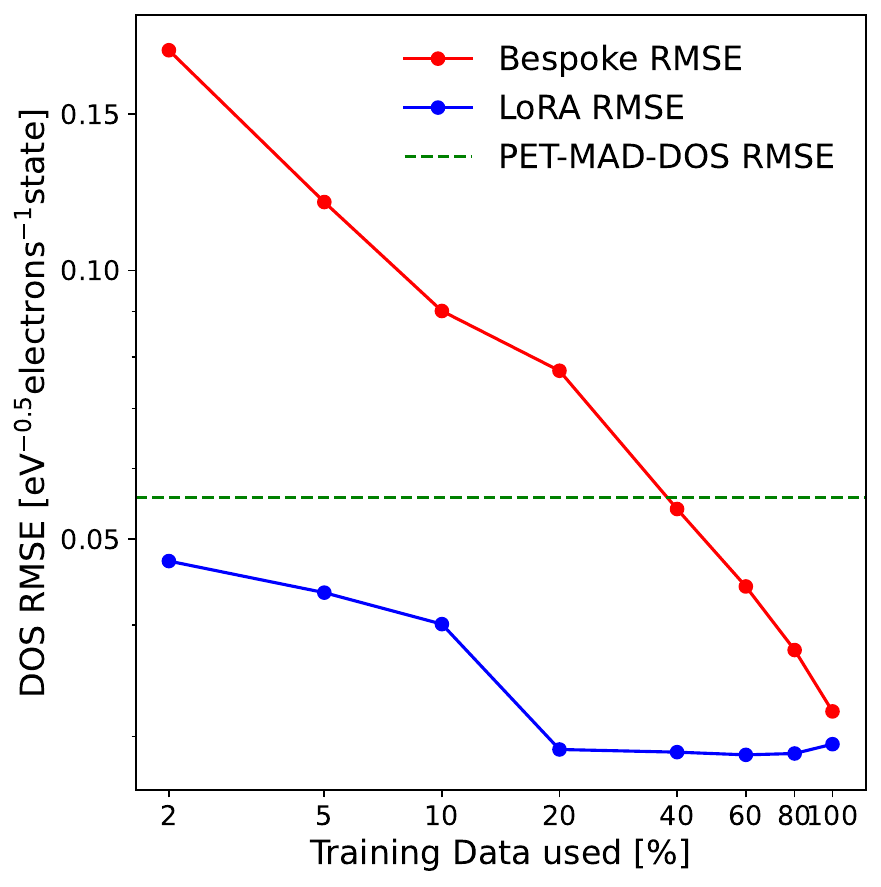}
    \caption{Learning curves for the HEA dataset, comparing the performance of the bespoke model and the LoRA fine-tuned model and that of the PET-MAD-DOS model. The amount of training data, randomly sampled from the HEA training set (1577 structures), is represented on the x-axis as a percentage and the Test DOS RMSE is represented on the y-axis.}
    \label{fig:HEA_LC}
\end{figure}

\autoref{fig:HEA_LC} shows the learning curves for the high entropy alloy (HEA) dataset. The behaviour is similar to that of Li$_{3}$PS$_{4}$. The bespoke test errors have yet to saturate while the LoRA models saturated at 20\% training data, showing that LoRA models require significantly less data than bespoke ones.

\clearpage

\section{Model predictions for finite-temperature material simulations}

Since the MD predictions in Figure 6 of the main text were truncated to highlight the most relevant sections of the DOS, this section presents a larger range of the prediction, omitting only the regions below the pseudo-core states where the DOS is zero and very high energies where the DOS are unreliable and cannot be compared meaningfully. The thermal-average DOS are computed simply as follows,
\begin{align}
    \mathrm{DOS_{average}}(E) = \frac{1}{N}\sum_A\mathrm{DOS_A}(E)
\end{align}
where N represents the number of structures in the trajectory and A represents the index of the structure.

\begin{figure*} [h]
    \centering
    \includegraphics[width=\textwidth]{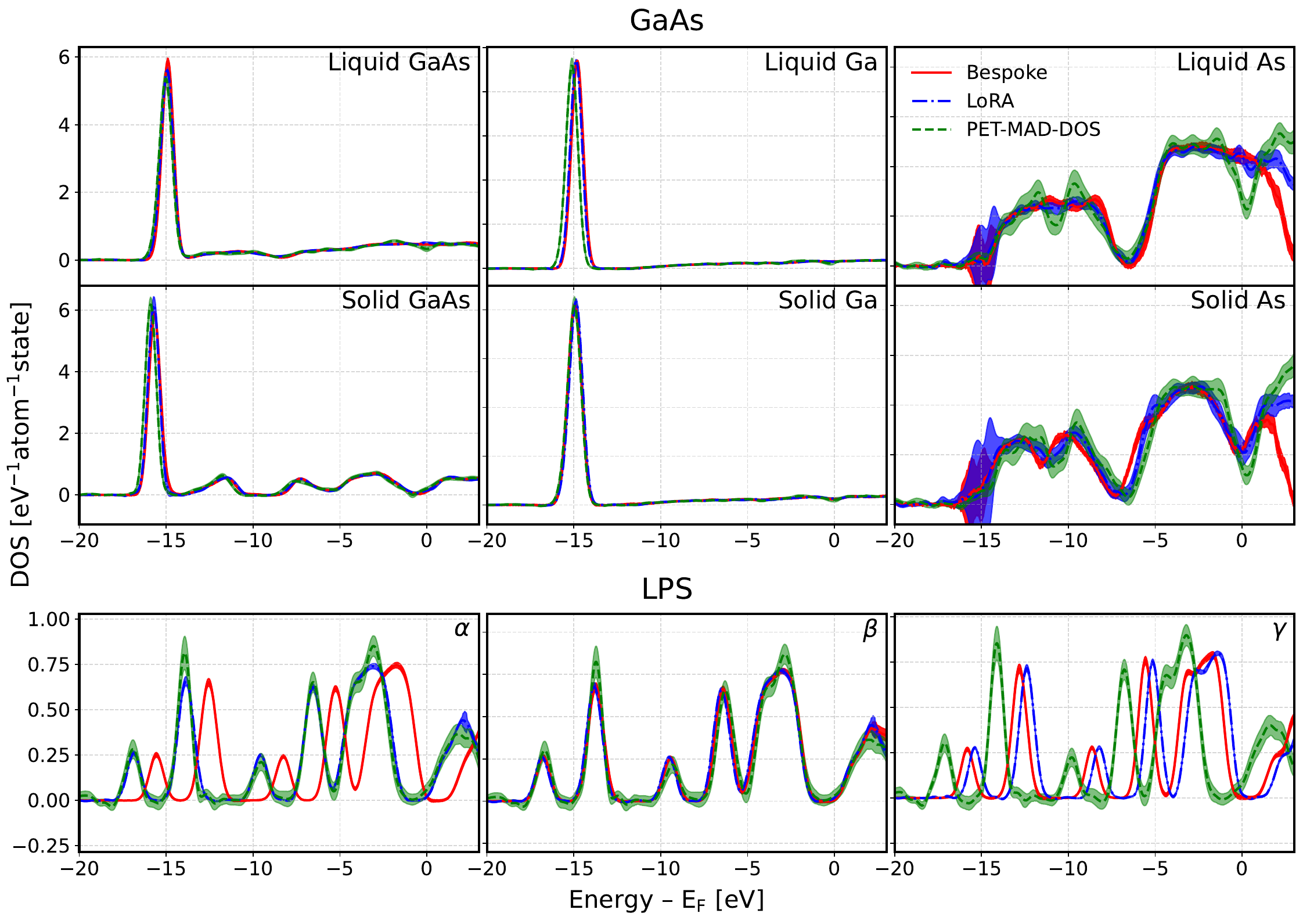}
    \caption{\label{fig:MD_predictions_full} Full DOS predictions of the MD trajectories of GaAs (top 2 rows) and LPS (bottom row) at different phases. The red solid lines represent the prediction of the bespoke model, the blue dash-dotted lines represent the prediction of the LoRA model, and the green dotted line represents the prediction of PET-MAD-DOS. The colored areas represent the uncertainty associated with the DOS predictions of the corresponding model, obtained by propagating the uncertainties from each individual snapshot in the MD trajectory. In this procedure, the thermal-average DOS is computed for each member in the calibrated last-layer prediction rigidity (LLPR) ensemble, and the standard deviation across the ensemble members is taken as the uncertainty. Each system's phase is labelled at the top right corner of each subplot. The MD trajectories are obtained using a bespoke PET-MAD model. The axis for all systems is truncated to remove high-energy regions where the predictions are unreliable and energy below the pseudo-core states where the DOS is zero. For all subplots, the DOS is normalized with respect to the number of atoms in the system and the energy reference is set to the Fermi level determined based on each respective DOS prediction.}
\end{figure*}

From \autoref{fig:MD_predictions_full}, we can see that although there are some deviations in the DOS profile for pseudo-core states, it did not impact the Fermi level determination significantly, as the DOS lines up relatively well across all 3 models. This can be seen more prominently in Fig. 3 of the main text. 

\begin{figure*}
    \centering
    \includegraphics[width=\textwidth]{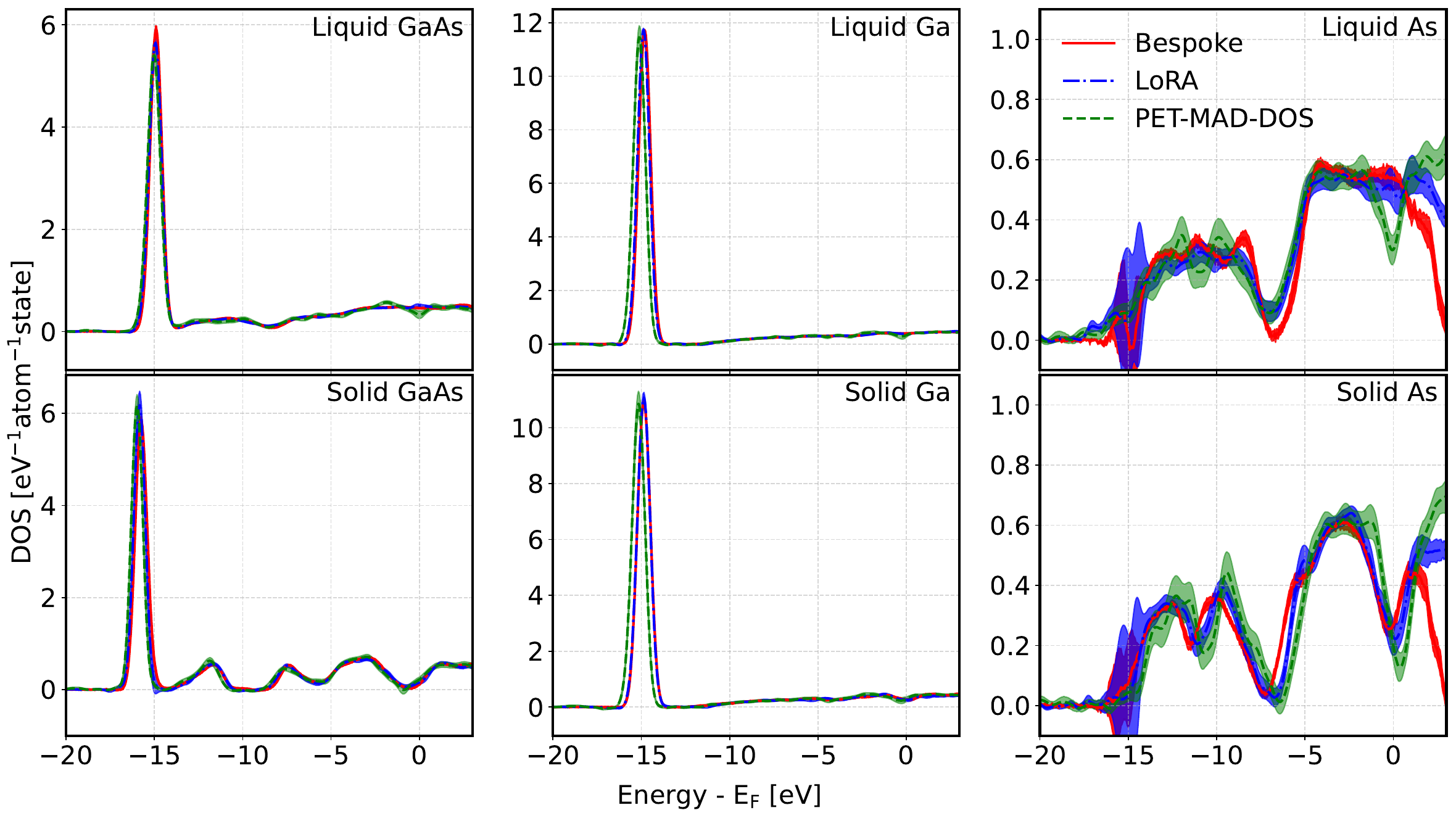}
    \caption{\label{fig:universal_GaAs_MD} Full DOS predictions of the MD trajectories of GaAs at different phases, with the MD trajectories obtained using the PET-MAD MLIP. The red solid lines represent the prediction of the bespoke model, the blue dash-dotted lines represent the prediction of the LoRA model, and the green dotted line represents the prediction of PET-MAD-DOS. The colored areas represent the uncertainty associated with the DOS predictions of the corresponding model, obtained by propagating the uncertainties from each individual snapshot in the MD trajectory. In this procedure, the thermal-average DOS is computed for each member in the calibrated last-layer prediction rigidity (LLPR) ensemble, and the standard deviation across the ensemble members is taken as the uncertainty. Each system's phase is labelled at the top right corner of each subplot. The axis for all systems is truncated to remove high-energy regions where the predictions are unreliable and energy below the pseudo-core states where the DOS is zero. For all subplots, the DOS is normalized with respect to the number of atoms in the system and the energy reference is set to the Fermi level determined based on each respective DOS prediction.}
\end{figure*}

Additionally, we have computed the same MD trajectories using the PET-MAD MLIP instead of the bespoke PET MLIPs. As both set of results are nearly identical, the thermal-average DOS from the bespoke PET MLIP was reported in the main text. Here, we present the thermal-average DOS from the PET-MAD MLIP as well in \autoref{fig:universal_GaAs_MD}. 

\clearpage

\section{Model Performance in the high-energy range}
\begin{figure} [h]
    \centering
    \includegraphics[width=0.5\textwidth]{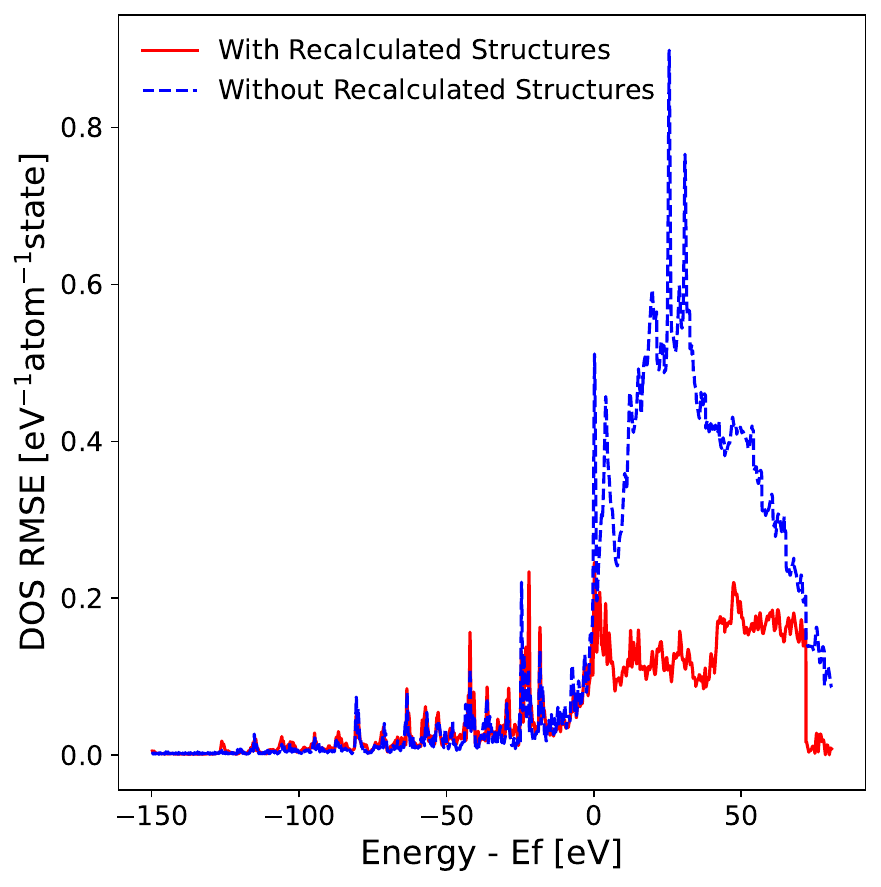}
    \caption{Figure comparing the RMSE of the predictions, at each energy channel, of a PET-MAD-DOS model trained on datasets with and without the recalculated structures in the dataset. The error is evaluated on the recalculated structures on the test set. The red line depicts the RMSE at every energy channel for the model trained on recalculated structures while the blue line depicts that of the model trained without recalculated structures. The error is computed by simply taking the RMSE, at each energy channel, between the prediction and target at the alignment that minimizes the metric in Eq. \eqref{ALoss} of the main text.}
    \label{fig:energywise_gradient}
\end{figure}

The model's performance at high-energy regions can be important in high temperature applications or in systems with large bandgaps, where the virtual states have high energies. To enhance model performance at high energies, a small subset (850 structures) has been recomputed with 4 times the number of valence bands. In \autoref{fig:energywise_gradient}, it can be observed that including the recalculated structures resulted in a significant decrease in the prediction errors in high-energy regions when evaluated on the recalculated structures in the test subset.  The errors begin to deviate significantly after the Fermi level of the structures, with the error of the model without recalculated structures far exceeding that of the model with recalculated structures.

\begin{figure*}
    \centering
    \includegraphics[width=\textwidth]{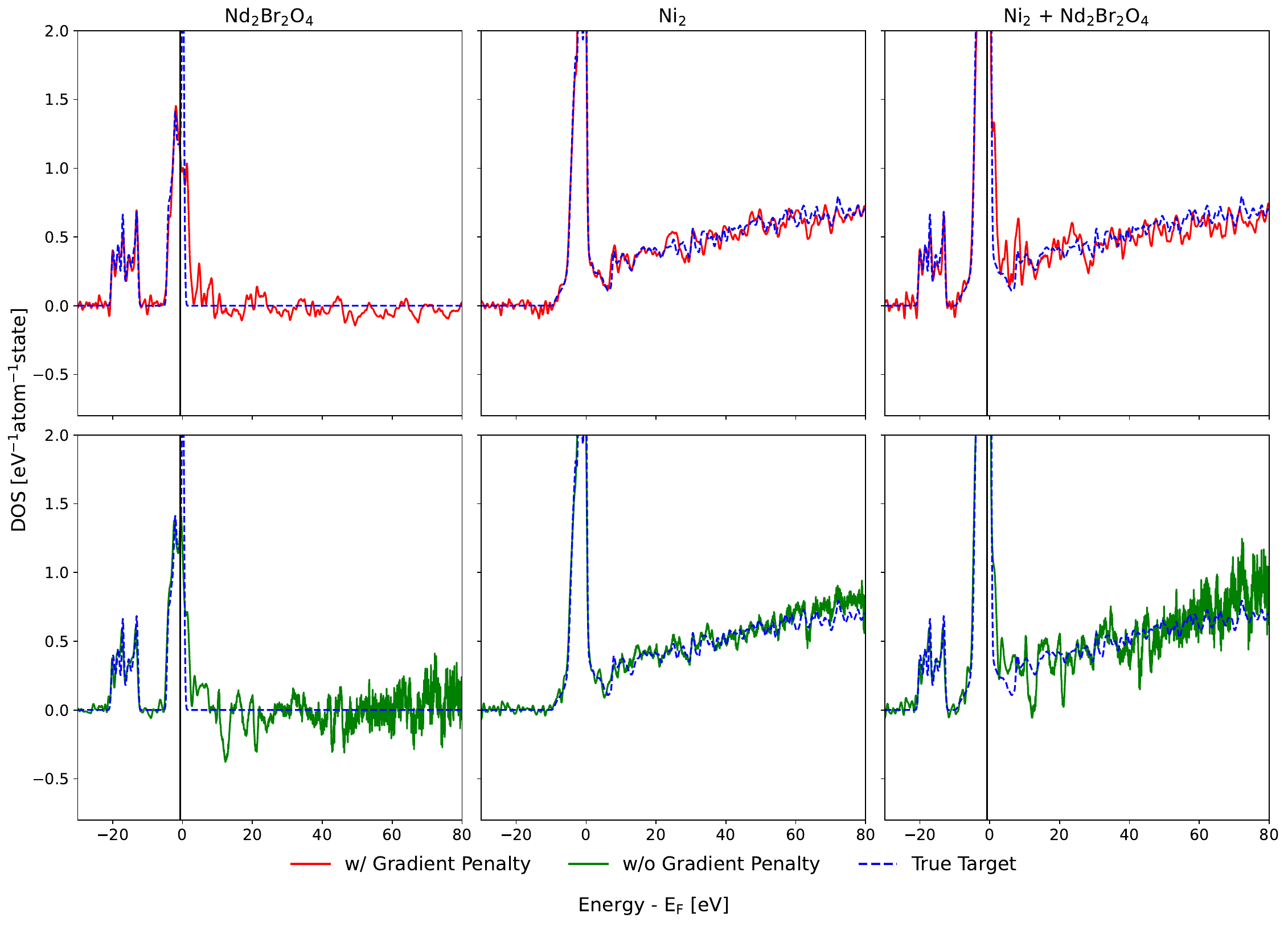}
    \caption{\label{fig:gradient_sample} Model predictions on a training structure with the lowest energy cutoff (\ce{Nd2Br2O4}) and highest energy cutoff (\ce{Ni2}). The \ce{Nd2Br2O4} belongs in the MC-2D subset while \ce{Ni2} belongs in the MC-3D subset. 
    The red line depicts the predictions from the model trained with gradient penalty while the green line depicts that of a model trained without the gradient penalty. The black vertical line denotes the energy cutoff $E_{max}$ of \ce{Nd2Br2O4} while the $E_{max}$ of \ce{Ni2} exceeds the prediction window and is not depicted. The true target for \ce{Nd2Br2O4} + \ce{Ni2} is computed by simply summing up the true target in the first 2 columns, hence the DOS at high energies do not include contributions from \ce{Nd2Br2O4}. The y-axis has been truncated to make the effects more prominent. The sudden drop in the DOS for \ce{Nd2Br2O4} arises due to the limited number of eigenvalues in the DFT calculation. As observed, the strong oscillations in the \ce{Nd2Br2O4} prediction of the model trained without gradient penalty contaminated the predictions of \ce{Ni2}, resulting in worse prediction quality in the combined system.}
\end{figure*}

Furthermore, the inclusion of the gradient penalty in the training loss function alleviates the issue of rapid oscillations in the predictions above the energy cutoff ($E_{max}$) due to lack of data. These oscillations can contaminate the predictions if the structure to be evaluated contains atomic environments from training structures that have very different $E_{max}$. We demonstrate this in \autoref{fig:gradient_sample}, where we combined the predictions of two training structures, one with low $E_{max}$ (\ce{Nd2Br2O4}) and one with high $E_{max}$ (\ce{Ni2}). The black vertical line denotes the $E_{max}$ of \ce{Nd2Br2O4}. Since the $E_{max}$ of \ce{Ni2} exceeds the prediction window, it is not shown in the plot. Despite both models performing well within the evaluation window (below $E_{max}$), the predictions of \ce{Nd2Br2O4} by the model trained without gradient penalty started to exhibit rapid oscillations roughly 40eV above the Fermi level while that of \ce{Ni2} did not exhibit those oscillations because its $E_{max}$ is above the prediction window. As a result, the prediction of the combined structure in the high-energy region is significantly worse for the model trained without gradient penalty due to oscillations from the structure with lower $E_{max}$ interfering with the predictions from the structure with higher $E_{max}$.

\clearpage

\section{Hyperparameters optimization}

To obtain the optimal model in terms of accuracy and computational speed, we performed a grid search over the hyperparameters on the Pareto front of the PET-MAD model. The summary of the hyperparameters are as follows:

\begin{description}
     \item[${R_{\mathrm{cut}}:}$] Cutoff radius defining the range for message passing between atoms
    \item[$N_\mathrm{GNN}:$] Number of message-passing layers
    \item[$N_\mathrm{trans}:$] Number of transformer layers in each message-passing layer
    \item[$d_\mathrm{PET}:$] Dimensionality of the messages
    \item [$N_\mathrm{heads}:$] Number of heads in the multi-head attention layers
\end{description}

\begin{figure}[h]
    \centering
    \includegraphics[width=0.5\textwidth]{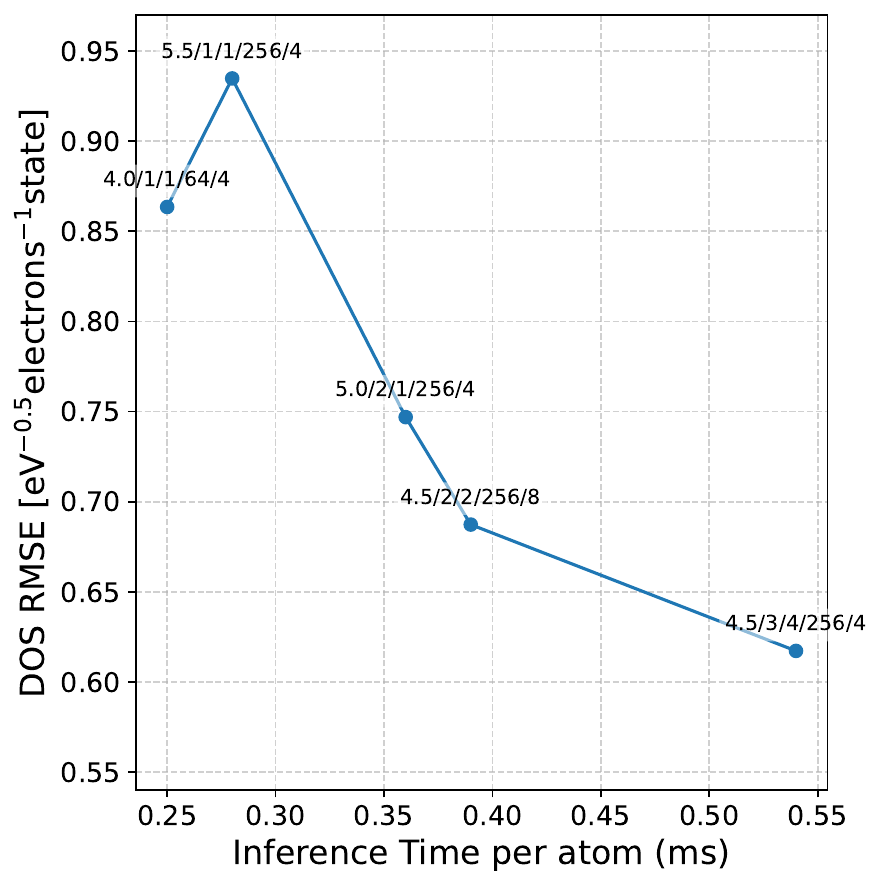}
    \caption{Performance of models trained on the hyperparameters that lie on the pareto front of PET-MAD. The x-axis represents the inference time per atom, measured on a single NVIDIA H100 GPU with a batch size of 1. The y-axis denotes the root mean square error (RMSE) on the DOS on the validation set.}
    \label{fig:pareto-frontier}
\end{figure}

The hyperparameters that lie on the pareto front of the PET-MAD model, using the notation [$R_{cut}/N_{GNN}/N_{trans}/d_{PET}/N_{heads}$], are [$4.0/1/1/64/4$], [$5.5/1/1/256/4$], [$5.0/2/1/256/4$], [$4.5/2/2/256/8$], [$4.5/3/4/256/4$]. For each set of hyperparameters, a separate training was performed. Model accuracy was evaluated on the validation set and the model inference time was measured using a single NVIDIA H100 GPU with a batch size of 1. The results are shown in \autoref{fig:pareto-frontier}. Based on the results obtained, the optimal hyperparameters were determined to be [$4.5/2/2/256/8$].

\clearpage
\section{Performance of Fermi level model}

\begin{figure}[!h]
    \centering
    \includegraphics[width=0.45\textwidth]{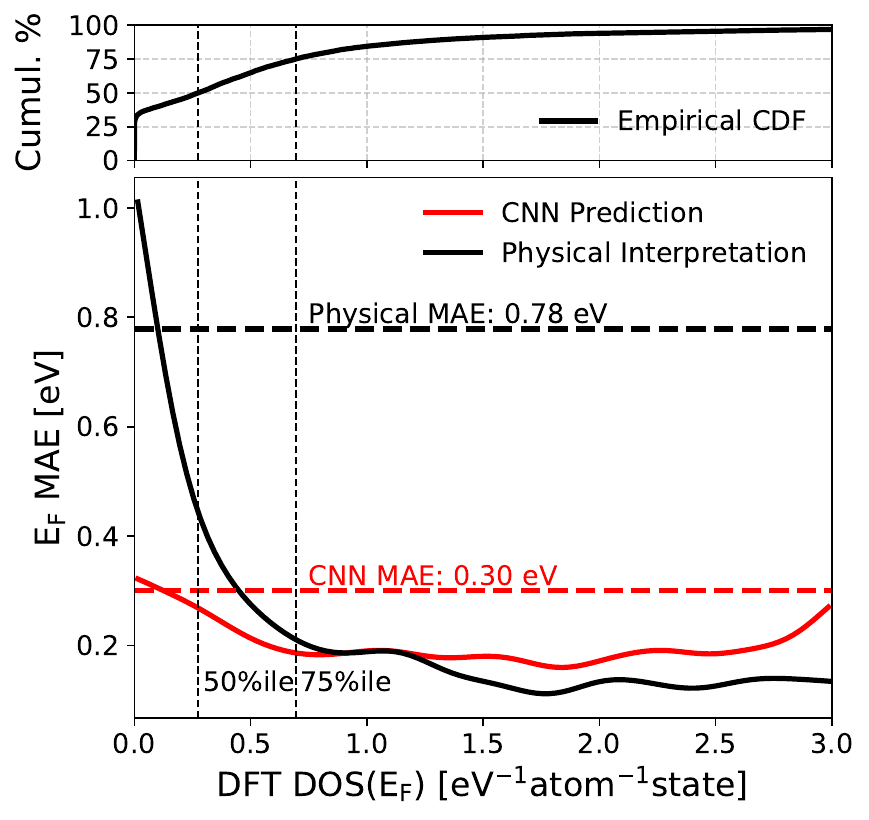}
    \caption{\label{fig:EfDOSEf}
    Variability of the Fermi level errors with the true DOS at the Fermi level, DOS($\mathrm{E_F}$), of the system. The two lines in the bottom subplot represent the mean absolute error (MAE) when obtaining the Fermi level by physical interpretation (black) or a convolutional neural network (CNN) (red). The x axis represents the DOS($\mathrm{E_F}$) of the system, as obtained from DFT calculations. The upper subplot contains the cumulative distribution (CDF) of DOS($\mathrm{E_F}$), expressed as a percentage of the test subset.
    }
\end{figure}

\autoref{fig:EfDOSEf} compares the performance of a convolutional neural network (CNN) model and the physical interpretation of the raw PET-MAD-DOS prediction for the purposes of determining the Fermi level. As observed, using CNNs is most useful when the DOS at the Fermi level is small, in which case integration errors would result in big shifts of the Fermi level. The majority of the MAD dataset (around 85\%) falls in the regime where using CNNs is beneficial, making them a better choice overall. However, one could come up with a threshold DOS($\mathrm{E_F}$) to switch to direct physical interpretation for the Fermi level computation.

\clearpage
\section{Fine-tuning accuracies}\label{sec:fine-tuning-accuracies}

For each simulation case presented in this work we trained a bespoke PET model from scratch, and compared it against the LoRA-fine-tuned version. While being equally accurate in predicting observables, the fine-tuned model retains a certain degree of accuracy on the base MAD dataset, which can be beneficial in certain computational setups. In \autoref{tab:fine-tuning-models-mad-accuracies}, we list the root mean square errors of each fine-tuned model in predicting the DOS on the base MAD test set. 

\begin{table}[h]
\renewcommand{\arraystretch}{1.2}
\centering
\begin{tabular}{lc}
\hline
\multicolumn{2}{c}{RMSE on MAD Test subset [$\mathrm{eV^{-0.5}electrons^{-1}state}$]} \\
\hline
LoRA Model & DOS RMSE \\
\hline
GaAs     & 0.075 \\
LPS     &  0.080\\
HEA   & 0.089\\
\hline
\hline
PET-MAD-DOS & 0.073\\
\hline

\end{tabular}
\caption{\label{tab:fine-tuning-models-mad-accuracies}
    DOS RMSE of the LoRA-fine-tuned models on the MAD test set. The test error of PET-MAD-DOS was also included for reference.}
\end{table}

\clearpage

\section{Performance of Uncertainty Quantification (UQ) Module}

\begin{figure}[h]
    \centering
    \includegraphics[width=0.9\textwidth]{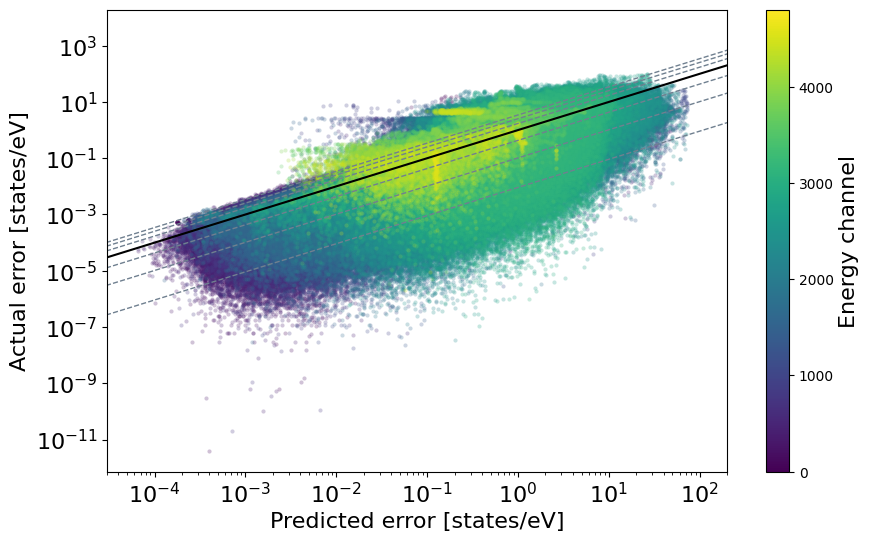}
    \caption{Parity plot of actual absolute error versus the estimated error from the LLPR-ensemble UQ module, presented in a log-log scale. The black dotted line delineates $y=x$. Each point corresponds to a prediction made for a test set structure for a given energy channel of PET-MAD-DOS. The grey dashed lines correspond to the isolines that are spaced $\sigma$ apart. The predicted uncertainties tell us that 68\% of the predictions should fall between the first set of isolines, then 95\% and 99\% for the two subsequent sets. The different energy channels are colored according to their channel index, with the lower indices corresponding to the lower energy regime of the DOS and vice versa.}
    \label{fig:UQ_Parityplot}
\end{figure}

The instantiation and calibration of the last-layer prediction rigidity (LLPR)-based UQ module was done as described in the main text. In calibrating the LLPR ensemble for the DOS models, the training set and validation set used in the training of the original model were equivalently employed. To align with the post hoc UQ calibration nature (i.e., to preserve the original model predictions), all model weights except for the last linear weights of the LLPR ensemble members were fixed during calibration. The calibration was performed globally with a single loss function that accumulates the error from all energy channels. Results in Figure~\ref{fig:UQ_Parityplot} show that this global calibration has been performed successfully, with most of the data point falling within the $3\sigma$ isolines. In general, small errors are observed for the earlier energy channels where the predictions are expected to be mostly zero, and higher errors in the energy channels in the latter energy channels. We note the existence of certain energy channels where the error distribution becomes complex for the following reason: for some structures, a peak exists in the DOS and the model must predict the nonzero peak, whereas for other structures, the DOS is supposed to be zero and hence the prediction must also be zero. This is especially prominent for the peaks corresponding to the core states of different elements. The calibrated uncertainties are still reasonable in these regimes, given that most of the data points still fall within the $3\sigma$ isolines. At the same time, however, we suspect that high errors committed during this complex prediction task may drive the rest of the uncertainties for the corresponding energy channels to the overestimation regime, whilst still leaving non-negligible number of points in the opposite regime where the errors are underestimated.

\clearpage
\begin{figure}[h]
    \centering
    \includegraphics[width=0.5\textwidth]{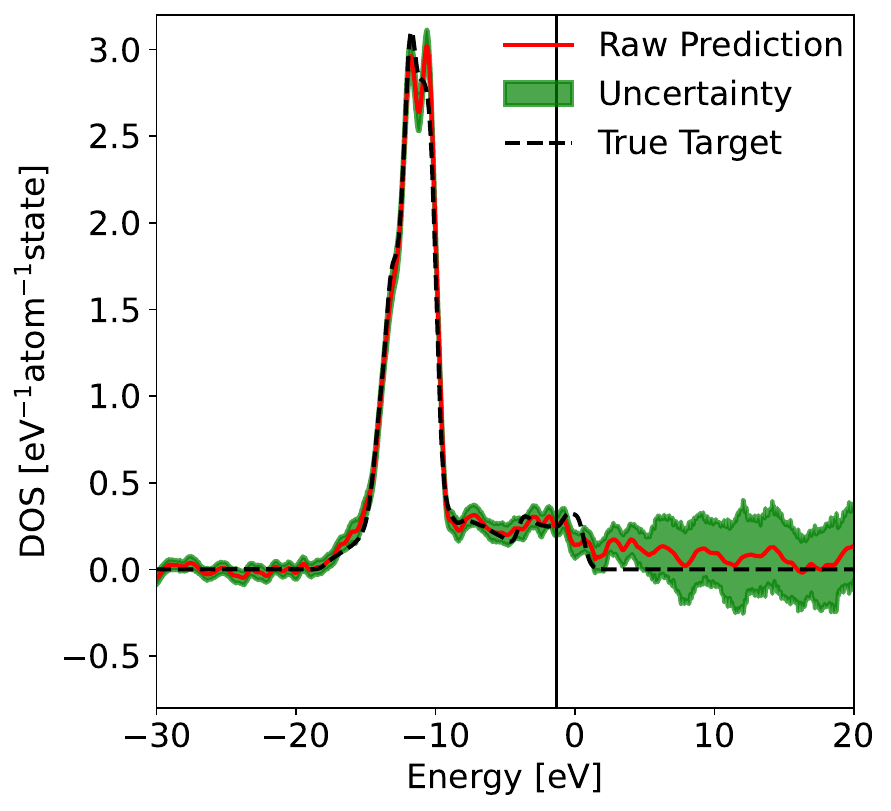}
    \caption{Demonstration of the UQ module on a sample test structure in determining the energy range where the model is extrapolating. The raw prediction is represented by the solid red line, and the true DOS target is represented by the dashed black line. The green area represents the uncertainty of the model, defined as the standard deviation of the calibrated LLPR ensemble. The vertical black line is the $\mathrm{E_max}$ of the structure, representing the energy cutoff of the DFT calculation.}
    \label{fig:UQ_extrapolation}
\end{figure}
In addition, the UQ module also accurately encapsulates the model's uncertainty at high energy channels. To tackle the low number of bands and wide range of eigenvalues in the dataset, the fitting of the model and ensemble uses a loss function with an adaptive window. As a result, most structures are not fit on the high energy channels of PET-MAD-DOS. As seen in \autoref{fig:UQ_extrapolation}, the UQ module reflects this behaviour well, manifesting as a spike in uncertainties past $\mathrm{E_max}$, where the model is fit on insufficient data.

\clearpage
\end{document}